\documentclass[iicol]{sn-jnl}% Default with double column layout

\usepackage{natbib}
\usepackage{graphicx}%
\usepackage{multirow}%
\usepackage{amsmath,amssymb,amsfonts}%
\usepackage{amsthm}%
\usepackage{mathrsfs}%
\usepackage[title]{appendix}%
\usepackage{xcolor}%
\usepackage{textcomp}%
\usepackage{manyfoot}%
\usepackage{booktabs}%
\usepackage{algorithm}%
\usepackage{algorithmicx}%
\usepackage{algpseudocode}%
\usepackage{listings}%
\usepackage{array}
\usepackage{float}
\usepackage[utf8]{inputenc}
\usepackage{collcell}
\usepackage{verbatim}
\usepackage{subfig}
\usepackage{lipsum, babel}
\usepackage{tabularx}
\usepackage{multicol}
\usepackage{appendix}
\usepackage{mathtools}
\usepackage{adjustbox} % Add this line to include the adjustbox package
\begin{document}

\title[Article Title]{Variable Chaplygin Gas: Constraining parameters using FRBs}


\author*[1]{\fnm{Geetanjali } \sur{Sethi}}\email{getsethi@ststephens.edu}
\equalcont{These authors contributed equally to this work.}
\author[1]{\fnm{Udish } \sur{Sharma}}\email{udishsharma1@gmail.com}
\equalcont{These authors contributed equally to this work.}

\author[1]{\fnm{Nadia} \sur{Makhijani}}\email{nadiamakhijani@gmail.com}
\equalcont{These authors contributed equally to this work.}

\affil[1]{\orgdiv{Department of Physics},\orgname{St Stephen's College},\orgaddress{\postcode{110007}, \state{Delhi},\country{India}}}


\abstract{We investigate cosmological constraints on the Variable Chaplygin gas model parameters with  latest observational data of the Fast Radio Bursts and compare the results  with previous constraints obtained using SNe Ia (Pantheon+SHOES), Gamma Ray Bursts, Baryon Acoustic Oscillations and Hubble parameter observational data. The Variable Chaplygin gas model is shown to be compatible with these datasets.
We have obtained tighter constraints on model parameters $B_s$ and $n$, using the FRB data set. By using the Markov chain Monte Carlo (MCMC) method  we obtain, $B_s$=$0.18\pm 0.10$ , $n=1.10\pm 1.15$  and $H_0$= $70.46\pm 0.66$ with the SNe Ia data set,  $B_s$= $0.09\pm0.06$ , $n= 0.44\pm 0.89 $  and $H_0=70.57\pm0.64 $ with the FRB data set, $B_s$=$0.16\pm 0.11$ , $n=1.06\pm 1.25$  and $H_0$= $70.37\pm 0.65$ with the BAO data set, $B_s$=$0.05\pm 0.000$ , $n=1.46\pm 0.23$  and $H_0$= $70.21\pm 0.57$ with the H(z) data set and $B_s$=$0.20\pm 0.11$ , $n=1.25\pm 1.17$  and $H_0$= $70.37\pm 0.64$ with the GRB data set . A good agreement for  $H_0$ is observed from these data sets.
}


\maketitle
\section{Introduction}\label{section 1}

Over the years, through various developments in the field of observational astronomy and astrophysics, cosmologists have been able to provide, with constant improvements, many frameworks that explain the evolution of our Universe on great timescales. 
The homogeneous and isotropic nature of the 
Universe on large length scales($\approx 100 Mpc$) has led to the development of the FLRW Metric:
\begin{equation*}
    ds^2 = c^2 dt^2 - a^2 (t)\left[ \frac{dr^2}{1-kr^2} + r^2(d\theta^2 + \sin^2 \theta d\phi^2 )  \right]
\end{equation*}
where k=$\pm 1,0$ defines the curvature of our Universe. Observationally, it has been found that a $k=0$ universe is viable.
This however has been proven to be inadequate. Through advancements in Astronomy with the measurements of galactic rotation curves which led to the \textit{missing mass problem} and later through the measurements of SNe Ia light curves (~\cite{Perlmutter_1999} and ~\cite{Filippenko_1998}) which for the first time  established the long theorised expanding nature of the Universe through observations, shortcomings in just the FLRW metric are present. \\

This led to the emergence of the $\Lambda- CDM$ model of Cosmology. This model introduced the idea of the presence of dark matter and dark energy, two unique quantities with properties wildly different from Baryonic matter. Dark matter is theorised to exist in Halos surrounding galaxies and is said to undergo only gravitational interactions. Dark energy has more varied interpretations but the idea that it exerts a negative pressure driving the accelerated expansion of the universe has been established. The $\Lambda- CDM $ model, also now known as the general cosmological model, has been able to predict accurately many significant observations. The model has been successful in explaining the Cosmic microwave background and its temperature fluctuations, the polarisation, and the abundance of light-matter in our Universe. It also explains the phenomena of gravitational lensing with the distribution that the Universe is composed of $68 \%  $ dark energy, $27\% $ dark matter, and $5\%$ ordinary matter. It explains the uneven distribution of lensing and greater lensing around galactic clusters due to the higher presence of matter. The model has also been able to explain the filament-like structures(\cite{1986ApJ...303...25T} )that exist in regards to the existence of galactic clusters. \\
Another important success that has been associated with the General Cosmological model is its ability to explain the redshift-distance relation. Redshift refers to the displacement of the spectrum of an astronomical object towards longer wavelengths. The model takes into account the contributions from baryonic matter, dark matter, and dark energy and successfully puts into picture the redshift of various astronomical sources moving away from us. This is consistent with the general overview of an accelerated expanding Universe. 
Despite of the success of the $\Lambda-CDM $ model making key predictions that fall well in line with observed phenomena, there have been a lot of shortcomings in the model. \\
The fine-tuning problem refers to the model's requirement of extremely tight constraints on values of many key parameters. If these constraints were not adhered to, the Universe would not exist the way we know it to be. The ratio of the gravitational force and the electromagnetic force is one such constraint that allows for the formation of the stars and their long lifetimes. A deviation from the currently established ratio would result in a Universe just filled with hot dense gas. The Hydrogen-Helium fusion reactions in the core of the stars can only be sustained by a very tight bound on the mass of a proton. A lighter proton would result in too rapid of a reaction leading the stars to burn out in short periods while a heavier proton would lead to no reaction at all. The evolution of the Universe also as we know it requires an extremely fine-tuned value of the Hubble parameter and the debate about the coincidence of the occurrence of these precise values which support the model and its predictions has been on for quite some time. Another significant objection to the model is Hubble tension(\cite{Thakur_2023}). Hubble tension refers to the discrepancy in the measurement of the Hubble constant from the two different methods of measurement. When calculated through observations of the Cosmic Microwave Background, the value of Hubble constant at the current epoch is found to be $H_0= 67 Km/s/Mpc$ and when measured through SNe Ia is found to be $H_0=73 Km/s/Mpc$. This is a difference of the order of 5 $\sigma$ that is very substantial and has pushed researchers forward to look for a unified theory that can explain both sets of observational data. Moreover, the rather mysterious nature of Dark Matter and Dark energy themselves and the fact that there has been no direct detection of dark matter further cast doubts on the viability of something like the $\Lambda-CDM $ model.
\\
\\
There have been alternates such as the Quintessence dominated models(\cite{PhysRevD.37.3406}) which propose the existence of a scalar field with negative pressure as a candidate for dark energy. These models have been able to explain the accelerated expansion of the universe and also do away from the existence of a cosmological constant $\Lambda$ which stays consistent over different epochs. Even though the quintessence dominated models describe in great effect the observations being made, they have their share of issues. Little to no concrete explanation is available for the scalar field which encompasses dark energy. The models also suffer from their own set of fine-tuning problems. \\
As an alternative hence, in this paper, we provide a study of one of the Chaplygin gas models. The generalisation to the original Chaplygin gas (\cite{chaplygin1904}) model by Sergei Chaplygin was introduced in 2003 (\cite{ocg}), these models(\cite{refId0})(\cite{PhysRevD.87.083503}) propose the existence of fluid with an exotic equation of state to explain the accelerated expansion of the Universe(\cite{Bamba2012}). A mathematical introduction to the model is provided in section (\ref{sec2}) followed by our analysis and estimation of the model parameters and the Hubble constant in the sections that follow.\\ 
For this analysis, we have used Fast radio bursts as a cosmic probe. FRBs are short-duration transient bursts that occur in the radio wavelengths. They have a well-established DM- redshift relation which allows the study of their properties in detail. FRBs have been shown to be good candidates for cosmological probes from the already done studies. (\cite{Liu2023-mk})(\cite{Yang2022})(\cite{10.1093/mnras/stad3708}) Along with FRBs we have also summarised and compared results from SNe Ia data, GRBs, Hubble parameter data and BAO. We have compared the parameter space with previously obtained results from all these probes \cite{Chraya_2023, aggarwal2019variable}.

\section{Variable Chaplygin Gas Model }\label{sec2}
The Chaplygin gas can be expressed by the equation of state $ P = \frac{-A}{\rho}$. Here A is a positive constant. 
The generalised Chaplygin gas model is characterised by the equation of state:
\begin{equation}\label{chaplygineq}
P_{ch}=-\bigg(\frac{A}{\rho_{ch}^{\alpha}}\bigg)
\end{equation}
where $\alpha$ is a constant such that $0< \alpha \leq 1$.
\\If we take the time component of the Energy-Momentum conservation equation, $T_{\mu} ^{\mu \nu} =0$, we obtain the continuity equation:
\begin{equation}
\frac{\partial \rho}{\partial t}+3 H(p+\rho)=0
\label{continuityeq}
\end{equation}

where H is the Hubble parameter, $H=\frac{\dot{a}}{a}$,  $\rho$ = total density of Universe. Using equation (\ref{continuityeq}), the evolution of energy density of the Chaplygin gas model is given as(\cite{bento}) :
\begin{equation}
\rho{}_{ch} = \bigg( A+\frac{B}{a^{3(1+\alpha)}} \bigg) ^ {\frac{1}{1+\alpha}}
\label{chaplyginrhoarelation}
\end{equation}
Here $a$ is the scale of the universe in its current epoch and B is a constant of integration. For the earlier epochs,$a<<1$. Taking this into account, equation (\ref{chaplyginrhoarelation}) transforms into $\rho \propto a^-3$. Thus in this case, the Chaplygin gas shows behavior like Cold Dark Matter(CDM). In the late epochs, when the Universe Scale factor $a>>1$, equation (\ref{chaplyginrhoarelation}) transforms to $p=-\rho= constant$. Hence, the behaviour of Chaplygin gas is like the cosmological constant at later times therefore implying the accelerated expansion of the Universe.
\\The Chaplygin gas equation of state results in a component that acts as non-relativistic matter at the early stage of the universe and acts as Cosmological constant equal to $ 8\pi GA^{\frac{1}{1+a}}$ at later stages.
The Chaplygin gas-based models have not withered down in terms of interest among researchers for the promise they have shown in the past. \\
However, the shortcoming with the Chaplygin gas model is it produces oscillations or an exponential blowup of matter power spectrums which are inconsistent with observations (\cite{sandvik}). To avoid this instability, the combined effect of shear and rotation can be taken into consideration. This slows down the collapse with respect to the simple spherical collapse model(\cite{PhysRevD.87.043527}). Hence ,Variable Chaplygin Gas(VCG) has been proposed as a modification (\cite{Guo2007-tu},\cite{Sethi_2006}). In the VCG model, the equation of state for the Chaplygin gas across the epochs is free-flowing. The VCG model is dynamically stable as discussed in one of the sections below. The model is also a good fit for different Cosmological tests including CMBR (\cite{makhathini_13}). The VCG model comes from the dynamics of a generalised d-brane in a (d+1,1) space-time and can be described by a complex scalar field $\phi$. The action  can be expressed in the form of a generalised Born-Infield action(\cite{bento}). Considering a Born-Infield Lagrangian (\cite{sen2002tachyon})
\\
\begin{equation}
\mathcal{L}_{\mathrm{BI}}=V(\phi) \sqrt{1+g^{\mu \nu} \partial_{\mu} \phi \partial_{\nu} \phi}
\end{equation}
Here $V(\phi)$ is the scalar potential. When considering a spatially flat FLRW universe, the energy density and pressure can be defined as: $\ \rho=V(\phi)\left(1-\dot{\phi}^{2}\right)^{-1 / 2}$ and $P=-V(\phi)\left(1-\dot{\phi}^{2}\right)^{1 / 2}$. Correspondingly, the equation of state for Chaplygin gas can be written as:
\begin{equation}
P=-\frac{V^{2}(\phi)}{\rho}
\label{eos_eqn}
\end{equation}
The self interaction potential can be rewritten as a function of the scale factor: $V^2 (\phi)=A(a) $. The equation of state for VCG thus becomes:
\begin{equation}
P{}_{ch}=-\frac{A(a)}{\rho_{ch}}
\label{eq:equationofstate}
\end{equation}
We can define $A(a)=A_{0}a^{-n}$ \label{scale_fac} which is a positive function of the cosmological scale factor $a$ and $A_0$ and $n$ are constants. When we use the energy conservation equation (\ref{eq:equationofstate}) in a flat Friedmann-Robertson-Walker Universe, the evolution of the VCG density is given as :
\begin{equation}
\label{vcgevolution}
\rho{}_{ch}=\sqrt{\frac{6}{6-n}\frac{A{}_{0}}{a{}^{n}}+\frac{B}{a{}^{6}}}
\end{equation}
where B is the constant of integration. \\
Adding to this Einstein's Field equations  $G_{\alpha\beta}=8\pi G T_{\alpha\beta}$, and the FLRW metric, the Friedmann equations can be obtained:
\begin{equation}
\begin{aligned}
H^{2} &=\frac{8 \pi G}{3} \rho-\frac{k}{a^{2}} \\
\frac{\ddot{a}}{a} &=-\frac{4 \pi G}{3}(\rho+3 p)
\end{aligned}
\end{equation} 
Here $\rho$ is the baryonic matter density. Taking the dark energy component into consideration, the rate of expansion of the Universe can be expressed in terms of matter and radiation density $\rho$, curvature $k$, and the cosmological constant, $\Lambda$ as :
\begin{equation}
\label{freidmann}
H^{2} \equiv \bigg( \frac{\dot{a}}{a}\bigg)^{2} = \frac{8\pi G}{3}\rho-\frac{k}{a^{2}}+\frac{\Lambda}{3}
\end{equation}
Taking the critical density $\rho_c= \frac{3{H^{2}}}{8{\pi}G}$ and density parameter  $\Omega=\frac{\rho}{\rho_c}$, the first Friedmann equation can be written as :
\begin{equation}
    \Omega_{b}(a) + \Omega_{k}(a) =1
\end{equation}
We can extend this relation directly to other models with more components. When baryonic matter and radiation are taken into account, the Friedmann equation is written as :
\begin{equation}
        H^2=H_0^{2} (\Omega_{m,0} a^{-3}+ \Omega_{\gamma,0} a^{-4}+\Omega_{k,0} a^{-2})
    \end{equation}

where $\Omega_{m,0} + \Omega_{\gamma,0} + \Omega_{k,0}=1$. All other components can be added if their behavior with the scale factor is known. 
Reducing equation (\ref{freidmann}) to the case of a spatially flat Universe:
\begin{equation}
H{}^{2}=\frac{8\pi G}{3}\rho
\end{equation}
$H\equiv \dot{a}/a$ is the Hubble parameter. Hence, for the case of VCG model from equation (\ref{vcgevolution}) corresponding to the VCG density, the acceleration condition $\ddot{a} > 0$ is equivalent to:
\begin{equation}
\left(\frac{12-3n}{6-n}\right)a^{6-n}>\frac{B}{A_{0}}
\end{equation}
To allow for accelerated expansion of the Universe, the necessary condition is $n<4$. 
Taking $n=0$ restores the original Chaplygin gas behavior. Initially, the gas is like dust-like matter ($\rho_{ch} \propto a^-3$) and later acts as a cosmological constant ($p=-\rho= constant$). In the current state of VCG, the universe tends to be quintessence-dominated( $n>0$)(\cite{Hannestad_2002}\cite{PhysRevD.72.023504}) or phantom-dominated($n<0$)(\cite{PhysRevLett.91.071301}). The constant equation of state parameter $w\equiv -1+ \frac{n}{6}$. The first term of the right-hand side of Eq (\ref{vcgevolution}) is negligible initially so the equation can be written as $\rho \sim a^-3 $. This corresponds to a dust-like matter-dominated Universe. 
At present, the value of Energy density of VCG is: 
\begin{equation}
\rho_{cho}=\sqrt{\frac{6}{6-n}A_{0}+B}
\end{equation}
Here, $a_0 =1$. We can define a parameter $B_s$:
\begin{equation}
B_s=\frac{B}{6A_{0}/(6-n)+B}
\end{equation}
The energy density can be written in terms of $B_s$ and $n $ as :
\begin{equation}
\rho_{ch}(a)=\rho_{ch0}\left[\frac{B_s}{a^{6}}+\frac{1-B_s}{a^{n}}\right]^{1/2}
\label{eq:energydensity}
\end{equation}

The Friedmann equation in terms of the redshift z, taking $a= \frac{1}{1+z}$  and using equation (\ref{eq:energydensity}) becomes:
\begin{equation}
\begin{split}
 H^{2}=&\frac{8\pi G}{3}\bigg\{\rho_{r0}(1+z)^{4}+\rho_{b0}(1+z)^{3}+\\&\rho_{ch0}\Big[B_s(1+z)^{6}+(1-B_s)(1+z)^{n}\Big]^{1/2}\bigg\}
\label{eq:friedmanvcg}
\end{split}
\end{equation}

where $\rho_{r0}$ and $\rho_{b0}$ are the present values of energy densities of radiation and baryons, respectively. 
\footnote{\label{foot1}We have used the fact that for a flat Universe, $\Omega_{b0}+\Omega_{r0}+\Omega_{ch0}=1$, i.e the total matter density sums up to unity.}\\
Using (\ref{foot1})
\begin{equation}
\frac{\rho_{r0}}{\rho_{ch0}}=\frac{\Omega_{r0}}{\Omega_{ch0}}=\frac{\Omega_{r0}}{1-\Omega_{r0}-\Omega_{b0}}
\end{equation}
\\
and
\begin{equation}
\frac{\rho_{b0}}{\rho_{ch0}}=\frac{\Omega_{b0}}{\Omega_{ch0}}=\frac{\Omega_{b0}}{1-\Omega_{r0}-\Omega_{b0}},
\end{equation}
\\
Equation (\ref{eq:friedmanvcg}) becomes:
\begin{equation}
H^{2}=\Omega_{ch0}H_{0}^{2}a^{-4}X^{2}(a),
\label{FreidmannVCG}
\end{equation}
where
\begin{equation}
\label{X(a)}
\begin{split}
X^{2}(a)=&\frac{\Omega_{r0}}{1-\Omega_{r0}-\Omega_{b0}}+\\&\frac{\Omega_{b0}a}{1-\Omega_{r0}-\Omega_{b0}}+a^{4}\bigg(\frac{B_s}{a^{6}}+\frac{1-B_s}{a^{n}}\bigg)^{1/2}
\end{split}
\end{equation}

\section{Statistical Methodologies}
The Pantheon + SH0ES supernova data set (\cite{Scolnic_2022}) includes 1550 distinct type Ia supernovae and also includes observations from the host galaxies of 42 type Ia supernovae.The redshift of SNe Ia ranges from $z=0.001$ to $z=2.26$.This is the most recently compiled data set of SNe Ia. This data set provides us with the distance modulus of all the observation points along with their covariance matrix. The distance modulus ($\mu$) defined by the equation 
\[ \mu(z) = m_B (z) - M_B     \]
where $m_B$ and $M_B$ are the apparent and absolute magnitudes. It has been observed from previously done studies (\cite{2022JCAP...01..053K}) that absolute magnitude $M_B$ shows no evolution with redshift allowing us to make use of the distance modulus for the parameter estimation corresponding to the SNe Ia. There does exist a tension(\cite{2023arXiv230702434C}) which manifests in the value of $H_0$ upon calibration of the value of $M_B$ from CMB and SH0ES yet our analysis only focuses on the data calibrated using the latter. \\
For FRB, data of the localised FRBs has been compiled from the respective publications. Table(\ref{frb_table}) lists the data points used for our analysis.

\begin{table*}
\begin{tabular*}{\textwidth}{@{\extracolsep\fill}lccc}
\hline
NAME           & REDSHIFT (z) & $DM_{obs}$ ($pc$ $cm^{-3}$) & REFERENCES                                         \cr
\hline
rFRB 20121102A & 0.19273      & 557 $\pm$ 2               &  (~\cite{Tendulkar_2017}) \cr
FRB 20171020A  & 0.0087       & 114.1 $\pm$ 0.2           & (~\cite{Mahony_2018})  \cr
rFRB 20180301A & 0.3304       & 522$\pm$ 0.2              & (~\cite{Luo2020})   \cr
rFRB 20180916B & 0.0337       & 349.349 $\pm$ 0.005       &  (~\cite{Marcote2020})   \cr
rFRB 20180924C & 0.3214       & 361.42 $\pm$ 0.06         & (~\cite{Bannister_2019})\cr
FRB 20181030A  & 0.0039       & 103.396 $\pm$ 0.005       & (~\cite{Bhandari_2022})  \cr
FRB 20181112A  & 0.4755       & 589.27 $\pm$ 0.03         & (~\cite{Prochaska2019}) \cr
FRB 20190102C  & 0.2913       & 363.6 $\pm$ 0.3           &(~\cite{Macquart2020})  \cr
FRB 20190520B  & 0.241        & 1202 $\pm$ 10             & (~\cite{Niu_2022})      \cr
FRB 20190523A  & 0.66         & 760.8 $\pm$ 0.6           &(~\cite{Ravi2019}  )   \cr
FRB 20190608B  & 0.1178       & 338.7 $\pm$ 0.5           & (~\cite{Macquart2020})  \cr
FRB 20190611B  & 0.3778       & 321.4 $\pm$ 0.2           & (~\cite{Macquart2020} ) \cr
FRB 20190614D  & 0.6          & 959.2 $\pm$ 0.5           &(~\cite{Law_2020}) \cr
rFRB 20190711A & 0.522        & 593.1 $\pm$ 0.4           & (~\cite{Macquart2020} )\cr
FRB 20190714A  & 0.2365       & 504 $\pm$ 2               & (~\cite{Bhandari_2022})\cr
FRB 20191001A  & 0.234        & 506.92 $\pm$ 0.04         & (~\cite{Bhandari_2022})  \cr
FRB 20191228A  & 0.2432       & 297.5 $\pm$ 0.05          & (~\cite{Bhandari_2022})  \cr
rFRB 20200120E & 0.0008       & 88.96 $\pm$ 1.62          & (~\cite{Kirsten2022})  \cr
FRB 20200430A  & 0.1608       & 380.1 $\pm$ 0.4           &  (~\cite{Bhandari_2022})\cr
FRB 20200906A  & 0.3688       & 577.8 $\pm$ 0.02          & (~\cite{Bhandari_2022})\cr
rFRB 20201124A & 0.0979       & 413.52 $\pm$ 0.05         &(~\cite{Ravi2022})  \cr  
\hline
\end{tabular*}

\caption{Compiled data set of localised FRBs}
\label{frb_table} 
\end{table*}
The dispersive measure or DM of an FRB has contributions from a multitude of sources that can be encapsulated in the equation:
\begin{equation}
    \label{dm_igm_rel}
    DM_{obs} = DM_{MW}+ DM_{IGM}+ \frac{DM_{host} (z)}{1+z}
\end{equation}
Here $DM_{obs}$ is the observed DM. $DM_{MW}$ includes contributions from both the interstellar medium and the Milky Way halo, and can be written as: 
\begin{equation}
    DM_{MW}= DM_{Halo}+ DM_{ISM}
\end{equation}
The value of $DM_{halo}$ has been calculated to be between $50 \sim 80$ $ pc  \ cm^{-3}$ up to 200 KPc from the sun(\cite{10.1093/mnras/stz261} )Thus a normal distribution with a mean of $65 \ pc \ cm^{-3}$ with a standard deviation of $ 15 \ pc \ cm^{-3} $ has been assumed(\cite{Wu2022-zv}).

The NE2001(\cite{cordes2003ne2001i}) model takes into account galactic coordinates and has been used to derive the contribution from $DM_{ism}$. \\

 FRBs are classified into 3 categories based on the observational properties of the host galaxies. Category I includes repeating FRBs like FRB 121102 (\cite{Tendulkar_2017}) which was localised to a dwarf galaxy with stellar mass $\approx$ $4-7 X 10^7 M_{\odot} $.  FRBs like FRB 180916 (\cite{Macquart2020}) which originates from a host galaxy of stellar mass $10^{10}M_{\odot}$ are included in Category II. These are also repeating FRBs. The nonrepeating FRBs are a part of category III. The stellar masses of the origin galaxies of this type of FRBs range from $10^9 - 10^{10} M_{\odot}$. As part of \cite{Zhang2020-gb}, a log-normal function is used to fit the distribution of $DM_{host}$ obtained from the IllustrisTNG simulation(\cite{nelson2021illustristng}) which goes as :
\begin{equation}
P(x:,\mu,\sigma) = \frac{1}{x\sigma \sqrt{2 \pi}} exp \Big(-\frac{(lnx-\mu)^2}{2\sigma^2}  \Big)
\end{equation}
The evolution of median $DM_{host}$ with redshift can be given by the equation 
\begin{equation}
    DM_{host}= A(1+z)^{\alpha}
    \label{dm_host_equation}
\end{equation}
where $A$ and $\alpha$ are constants that differ for all the 3 categories of FRBs. The best-fit values of these constants for all the categories of FRBs, taken from \cite{Zhang2020-gb}, are in table \ref{constant_for_cateogry}.
 \begin{table}[h!]
    %\centering
    \begin{tabular}{ccc}
    \hline
        Category &  A & $\alpha$\\
        \hline\\
        
         I  & $34.72 ^{+17.77} _{-14.47}$   &  $1.08 ^{+0.87} _{-0.70}$ \\
         II  &   $ 96.22  ^{+50.10} _{-42.26}$   & $0.83 ^{+0.87} _{-0.58}$   \\
         III  &  $32.97 ^{+23.23} _{-17.65}$ &    $0.84 ^{+0.93} _{-0.60}$
         
    \end{tabular}
    \caption{Values and errors on $A$ and $\alpha$ from \cite{Zhang2020-gb}}
    \label{constant_for_cateogry}
\end{table}

% $DM_{igm}$ have been found using this and is what is used in the estimation of cosmological distances since it has a contribution purely from the intergalactic medium. \\
 Thus $DM_{igm}$ can thus be found using equation(\ref{dm_igm_rel}) and is what gets used in the cosmological study since the contribution is entirely from the intergalactic medium.\\
The uncertainty in $DM_{igm}$ is defined by the relation:
\begin{equation}
    \sigma_{IGM}(z)= \sqrt{ \sigma_{obs}(z) ^2 + \sigma_{MW} ^2 + \left( \frac{\sigma_{host}(z)}{1+z}  \right)^2 } 
\end{equation}
We have the $\sigma_{obs}$ for each observation and we can take the combined contribution of uncertainties in $\sigma_{halo}$ and $\sigma_{ism}$ to be $ 30 \ pc \ cm^{-3} $.\\
As for the uncertainty, $\sigma_{host}$,  we again refer to \cite{Zhang2020-gb} to define the value. The higher error limit and lower error limit on the constants $A$ and $a$ are used to define the uncertainty of $\sigma_{host}$. This differs as a function of redshift for all the 3 categories of FRBs since the bounds and values for the constants $A$ and $\alpha$ differ. \\

Having obtained the $DM_{igm}$ and $z$ data , we use the Macquart equation, 
(\cite{Macquart2020}):
\begin{equation}
    < DM_{igm}>= \frac{3C \Omega_b H_0}{8\pi G m_p} \int_0 ^z \frac{f_e(z')f_{igm}(z')(1+z') dz'}{E(z')}
    \label{macquart}
\end{equation}
This gives us the theoretical value of $DM_{igm}$.
We define $E(z)$ from equation(\ref{FreidmannVCG}) as 
\begin{equation}
    E(z)= \Omega_{ch0}^{1/2} a^{-2} X(a)
\end{equation}
Here $X(a)$ is from equation(\ref{X(a)})
.\\
$f_{igm} \approx 0.83$ is the fraction of the baryon mass in the IGM (\cite{2012ApJ...759...23S}) The electron fraction is $f_e (z)= Y_{H \chi_{e},H} (z) + \frac{1}{2}Y_{He \chi_{e},He} (z)   $ where $Y_H = 3/4$ and $Y_{He} = 1/4$ are the mass fractions of hydrogen and helium, respectively and 
$ \chi_{e,H}$ and $ \chi_{e,He}$ are the ionisation fractions of hydrogen and helium respectively. We set $f_e = 0.875$ since hydrogen and helim are both fully ionized when $z<3$ (\cite{doi:10.1146/annurev.astro.44.051905.092514}). Treating $<DM_{igm}>$ obtained from equation(\ref{macquart}) as the theoretical value of $DM_{igm}$ and writing it as  $DM_{th}$.From the table (\ref{frb_table}) above, we have not made use of FRBs with $z<=0.01$ due to their low values of DM. 
\\
 The distance modulus and luminosity distance have shown explicit dependence on the Variable Chaplygin gas model. 

For SNe Ia, since we have the values of distance modulus$(\mu_{obs})$ and redshift available in the data set, we consider distance modulus defined by the equation :
\begin{equation}
    \label{mu_dl}
    \mu = 5log_{10}(\frac{d_L}{Mpc}) + 25
\end{equation}
Also, we define luminosity distance $d_L$ using the Friedmann equation in a flat universe by:
\begin{equation}
    d_L = c(1+z)\int_0 ^z \frac{dz' }{H(z)}
\end{equation}
Taking $H(z) $ from the equation (\ref{eq:friedmanvcg}).This gives the  definition of $d_L$ as :
\begin{equation}
    \label{dl_rel}
     d_L =\frac{ c(1+z)}{H_o}\int_0 ^z \frac{dz' }{\Omega^{1/2} _{ch0} X(z) }
\end{equation}
We substitute the  obtained value of $d_L$ in equation(\ref{mu_dl}) to find the theoretical value of the distance modulus, $\mu_{th}$.\\

In equation (\ref{dl_rel}), we fit for the parameters $H_0$, $n$ and $B_s$. $X(z)$ comes from equation (\ref{X(a)}) where $a$ can be written as $a=\frac{1}{1+z}$.
Also, from the Pantheon data set, SNe with $z<=0.01$ have not been considered. 
Similar kinds of analysis can also be done for $H(z)- z $ datasets, essentially observed values of the Hubble parameter as a function of redshift. We can write the theoretical value of $H(z)$ using equation(\ref{FreidmannVCG}) and define the likelihood. 

 Baryonic Acoustic Oscillations can also  be used as Standard Ruler. The BAO distance measurements were adapted from \cite{aggarwal2019variable} and the references therein. The relation between distance and redshift is given by
\begin{equation}
   d_z=\frac{r_s(z_{drag})}{D_V(z)} 
\end{equation}
where, the volume-averaged distance, given that $d_C(z)=d_L(z)/(1+z)$,
\begin{equation}
    D_V(z)=\left[\frac{czd^2_C(z)}{H(z)}\right]^{1/3}
\end{equation}
and the radius of the comoving sound horizon at the drag epoch $z_{drag}$, 
\begin{equation}
    r_s(z_{drag})=\frac{c}{\sqrt{3}} \int\limits_{z_{drag}}^{\infty}  \frac{dz}{H(z)\sqrt{1+(3\Omega_{b0}/4\Omega_{r0}(1+z)^{-1})}}
\end{equation}
Gamma Ray bursts(\cite{Liang2008}) can also be used as a cosmological probe(\cite{liu2022}). Model independent use of GRBs is fairly popular (\cite{gaoetal})(\cite{reffId0})(\cite{Liang2022ApJ}). The correlation established by the \textit{Dainotti relation} allows us to do the same. It links the time at the end of the plateau emission measured in the rest frame, $T_X ^*$ with the corresponding X-ray luminosity of the LC, $L_X$. (\cite{2012ApJ...759...23S}). When extending this to 3 dimensions, a prompt peak luminosity, $L_{peak}$ is added, and this is known as the 3D Dainotti equation. ( \cite{2016ApJ...825L..20D} \cite{2017ApJ...848...88D})

Referring to \cite{10.1093/mnras/stac2752}, we write the log luminosity distance as 
\begin{equation}
\begin{split}
& log_{10}(D_L) = -\frac{log_{10}F_X + log_{10}K_X}{2(1-b)} + \\ & \frac{b.(log_{10}F_{peak}+log_{10}K_{peak})}{2(1-b)} -\frac{(1-b)log_{10}(4\pi)+ C}{2(1-b) }  + \\ & \frac{a(log_{10}T_X ^*)}{2(1-b)}
\end{split}
\label{grb_lum}
\end{equation}
The Platinum Sample of GRBs is used for this analysis. (\cite{Dainotti2020},\cite{10.1093/mnras/stac517})

\subsection{Monte Carlo Likelihood Analysis}

 Markov chain Monte Carlo algorithms have been used to do parameter estimation. These algorithms start with a set of parameters and initial proposal functions for all those respective parameters. Newer values are sampled for the parameters from all over the parameter space in accordance with the probabilities defined by the proposal function about the previous value and are accepted if they result in a higher probability. This way, the chain infallibly moves towards the set of parameters with the highest probability as the acceptance rate for new proposed parameters decreases. The advantage of this approach is its efficiency: the time required to sample from a distribution increases roughly linearly with the number of parameters, rather than exponentially. This makes MCMC techniques particularly valuable for estimating cosmological parameters, where the dimensionality of the parameter space can be very high.
For  $i^{th}$ data-point, given a certain set of parameters,  we define its Gaussian likelihood as :
\begin{equation}
    P_i= \frac{1}{\sqrt{2 \pi}} exp\left(-\frac{(x_i-\mu_{i})^2}{2\sigma_{i} ^2}\right)
\end{equation}
In this equation, $\mu_i$ acts as the theoretical value corresponding to that data point while $\sigma_i$ is the error bar associated with it. The combined probability of the entire data set can hence be defined as :
\begin{equation}
    \label{prelik}
    P_{total} = \prod_{i=1} ^n P_i(x_i | k)
\end{equation}
Where $k $ represents the set of parameters being sampled for. To get the to ideal value of $P_{total}$, we intend to find the optimal value for the set of parameters $k$. 
\\
Writing equation(\ref{prelik}) with $log(P_{total})$ as $\Theta$, we get to the relation:
\begin{equation}
    %\label{lk_frb}
    \Theta(k) = -\frac{1}{2}  \sum_{i} \frac{( x_{th} ^i - x_{obs} ^i)^2}{2\sigma_{i}^2}
\end{equation}
For FRBs, we can write the same as :
\begin{equation}
    \label{lk_frb}
    \Theta(H_0,B_s,n) = -\frac{1}{2}  \sum_{i} \frac{( DM_{th} ^i - DM_{igm} ^i)^2}{2\sigma_{i}^2}
\end{equation}
\\
Doing a similar analysis, we can define the likelihood for the SNe Ia data set as :
\begin{equation}
    \label{lk_sup}
    \Theta(H_0,B_s,n) = -\frac{1}{2}(\mu_{th} - \mu_{obs} )C^{-1} (\mu_{th} - \mu_{obs} )
\end{equation}
Here $C$ refers to our covariance matrix that is obtained from the data set.
Similarly, for Baryon Acoustic Observations,the likelihood function is defined  for the data set used, except the WiggleZ survey(\cite{Blake_2011}),  as
\begin{equation}
    \label{bao_lik1}
    \Theta=-\frac{1}{2}\sum_i \left[\frac{d^{th}_z-d^{obs}_z(z,\textbf{k})}{\sigma}\right] ^2
\end{equation}
and for the WiggleZ survey, the likelihood is defined by using the correlation matrix
\begin{equation}
    \label{bao_lik2}
    \Theta_{WiggleZ}=-\frac{1}{2}[d^{th}_{z,i}-d^{obs}_z(\textbf{p})]^T C^{-1} [d^{th}_{z,i}-d^{obs}_z(\textbf{k})]
\end{equation}
\\

Likelihood for $H(z)$ vs $z$ data is written as 
\begin{equation}
    \label{hz_lik}
    \Theta(H_0,B_s,n) = -\frac{1}{2} \sum_{i} \frac{(H_{th}- H_{obs})^2}{2\sigma_i ^2}
\end{equation} 
where as mentioned before, $H_{th}$ comes from equation(\ref{FreidmannVCG})

For the observational value, we can refer to the table in \cite{Li_2023} and the references therein.\\
 The theoretical value of Luminosity for GRBs can be described in a way similar to SNe Ia using equation(\ref{mu_dl}) and equation(\ref{dl_rel}). The  value of $\mu_{obs}$ for GRBs from equation(\ref{grb_lum}) as  
\[
\mu_{obs}= 5log_{10}(D_L) + 25
\]
The Likelihood for GRBs is accordingly then defined as 
\begin{equation}
    \label{grb_lik}
    \Theta(H_0,n,B_s) = -\sum_{i} log(\sigma_{\mu obs}^i ) -\frac{1}{2} \sum \frac{\mu_{th}^i-\mu_{obs}^i}{\sigma_{\mu obs_i} ^2}
\end{equation}
$\sigma_{\mu obs}$ is a term defined which has contributions of errors of all the constituent terms of equation (\ref{grb_lum}).

For the actual sampling, Algorithms like Metropolis-Hastings and Gibbs sampling are employed. We make use of  Cobaya (\cite{Torrado_2021}\cite{2019ascl.soft10019T})( Code for Bayesian Analysis) which is a framework designed for finding the posterior distributions of a set of parameters given their priors and their limits specially optimised for Cosmology studies (\cite{10.1093/mnras/stac517})(\cite{10.1093/mnras/stad3708}). 
The Cobaya samplers which include MCMC sampler and Polychord Sampler allow for a more efficient and faster sampling process. The functionalities of these samplers include oversampling of fast parameters allowing more steps to be taken along the axis where conditional distributions are easier to navigate.
The Cobaya package allows for modifications to be made and allows to create a new likelihood function. As part of the analysis, we define a new likelihood function for each different dataset and use this likelihood in the parameter estimation. 
In our case, in sampling for the parameters $H_0$, $B_s$, and $n$, we aim to maximise the value of $\Theta$ with respect to the set of these 3 parameters.
We use our likelihoods as defined in equations (\ref{lk_frb},\ref{lk_sup}),(\ref{hz_lik},\ref{grb_lik}) and equations(\ref{bao_lik1},\ref{bao_lik2})for Supernova ,FRBs ,GRBs, Hubble parameter data and Baryon Acoustic Oscillations respectively.\\
%\vspace{10cm}
\section{Results and Analysis}\label{sec4}
 We have used a  Gaussian prior  for sampling of the parameters in both cases, the Pantheon data set and the FRB data set. The means of prior distributions for $H_0$, $B_s$ and $n$ are taken to be  $70.4$, $0.11$ and $1.2 $ respectively. These are the values from previous constraints obtained using the Pantheon data set and doing a $\chi^2$ analysis. The estimations for the parameters of the VCG model using Cobaya with the likelihoods defined above for various datasets are given in Fig (1) to Fig (5) along with the overlap of contours obtained by FRB with SNe Ia( Pantheon)(Fig (6)), Baryon Acoustic Oscillations (Fig (7)), Gamma ray Bursts (Fig (8)) and 'H(z) vs z' datasets (Fig(9)). (\cite{aggarwal2019variable}\cite{Chraya_2023})
%\vspace{.5cm}

%\vspace{-.2cm}
%%%%%%%%%%%%%%%%%%%%
\begin{comment}

\begin{figure}[H]
\begin{center}

\begin{tabular}{cccc}
\subfloat[Distributions for the parameters $B_s$, $n$ and $H_o$ for the FRB Data set using VCG model]{\includegraphics[width = 2.5in]{fin_frb.png}} \\

\subfloat[Distributions for the parameters $B_s$, $n$ and $H_o$ for the SNe Ia Data set using VCG model]{\includegraphics[width = 2.5in]{fin_pantheon.png}}\\
\subfloat[Distributions for the parameters $B_s$, $n$ and $H_o$ for the BAO Data set using VCG model]{\includegraphics[width = 2.5in]{fin_bao.png}} \\

\end{tabular}

\end{center}
\end{figure}
\end{comment}
\begin{figure}[H]
 
\includegraphics[width=2in]{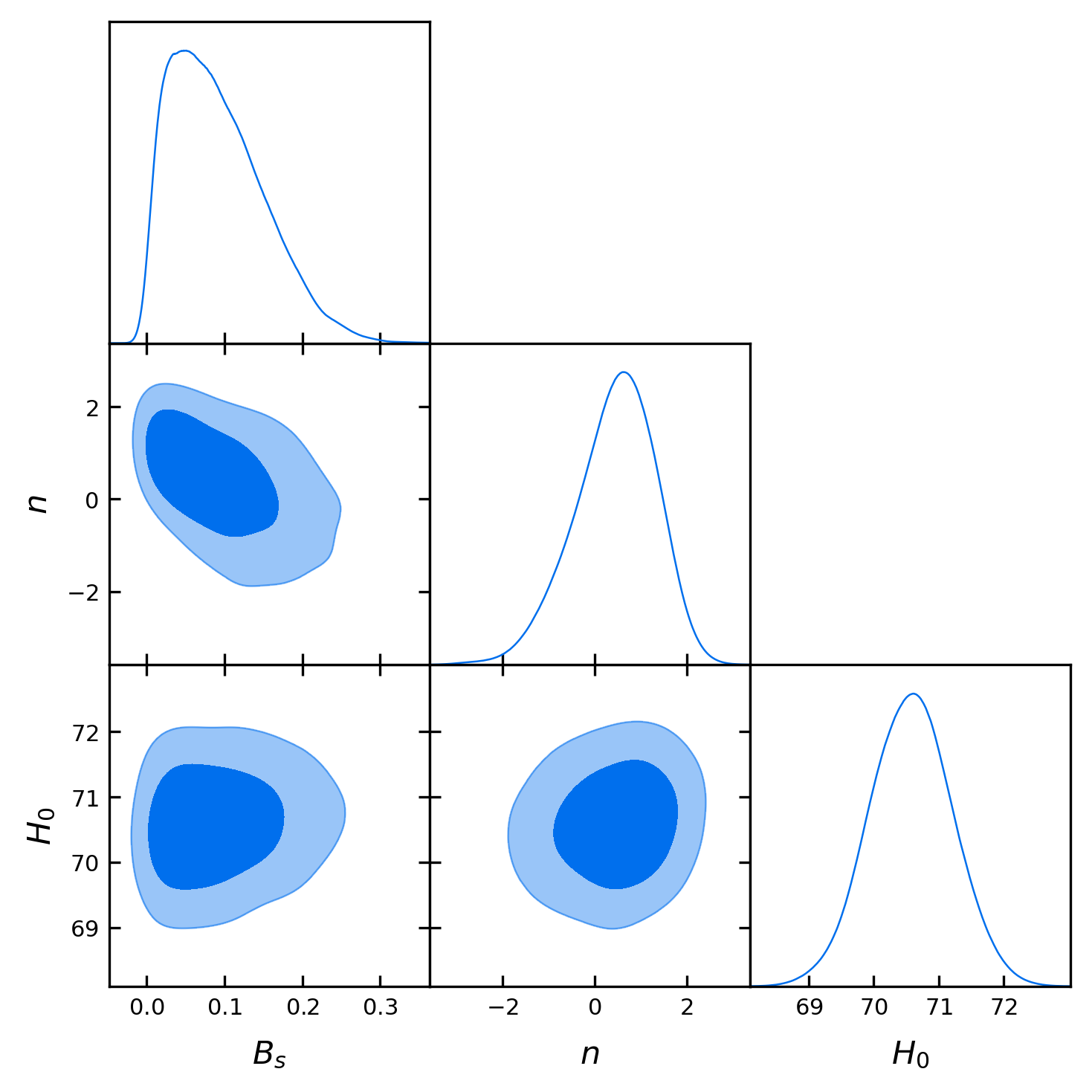}
\caption{Distribution and correlation of constraints obtained on model parameters and $H_0$ from the Fast radio Bursts data using the Macquart Equation. Results in table (\ref{final_table})}
\label{frb_result}
\end{figure}

\begin{figure}[H]
 
\includegraphics[width=2in]{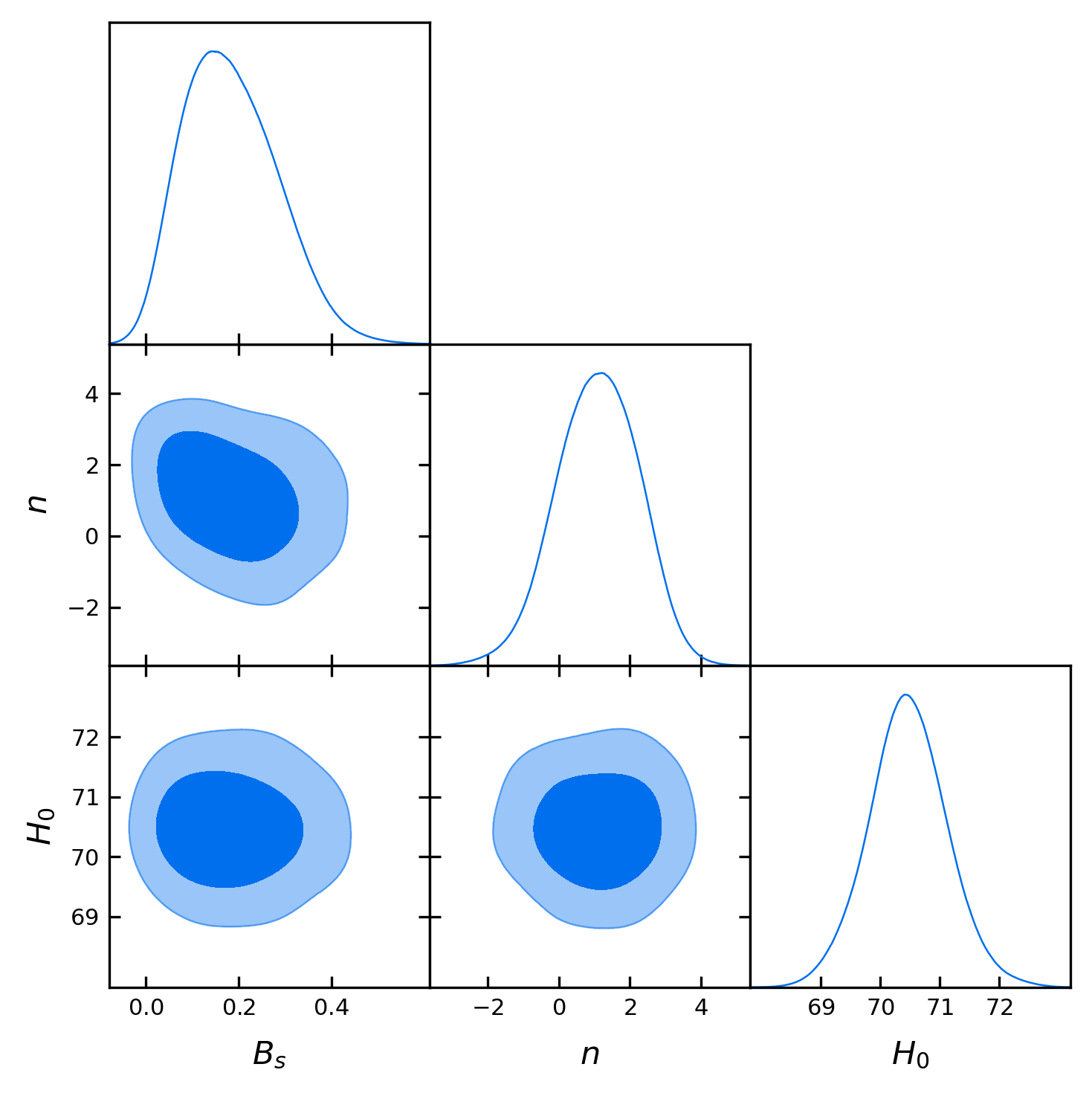}
\caption{Distribution and correlation of constraints obtained on model parameters and $H_0$ from the Pantheon+SH0ES data. Results in table (\ref{final_table})}
\label{sne_result}
\end{figure}

\begin{figure}[H]
 
\includegraphics[width=2in]{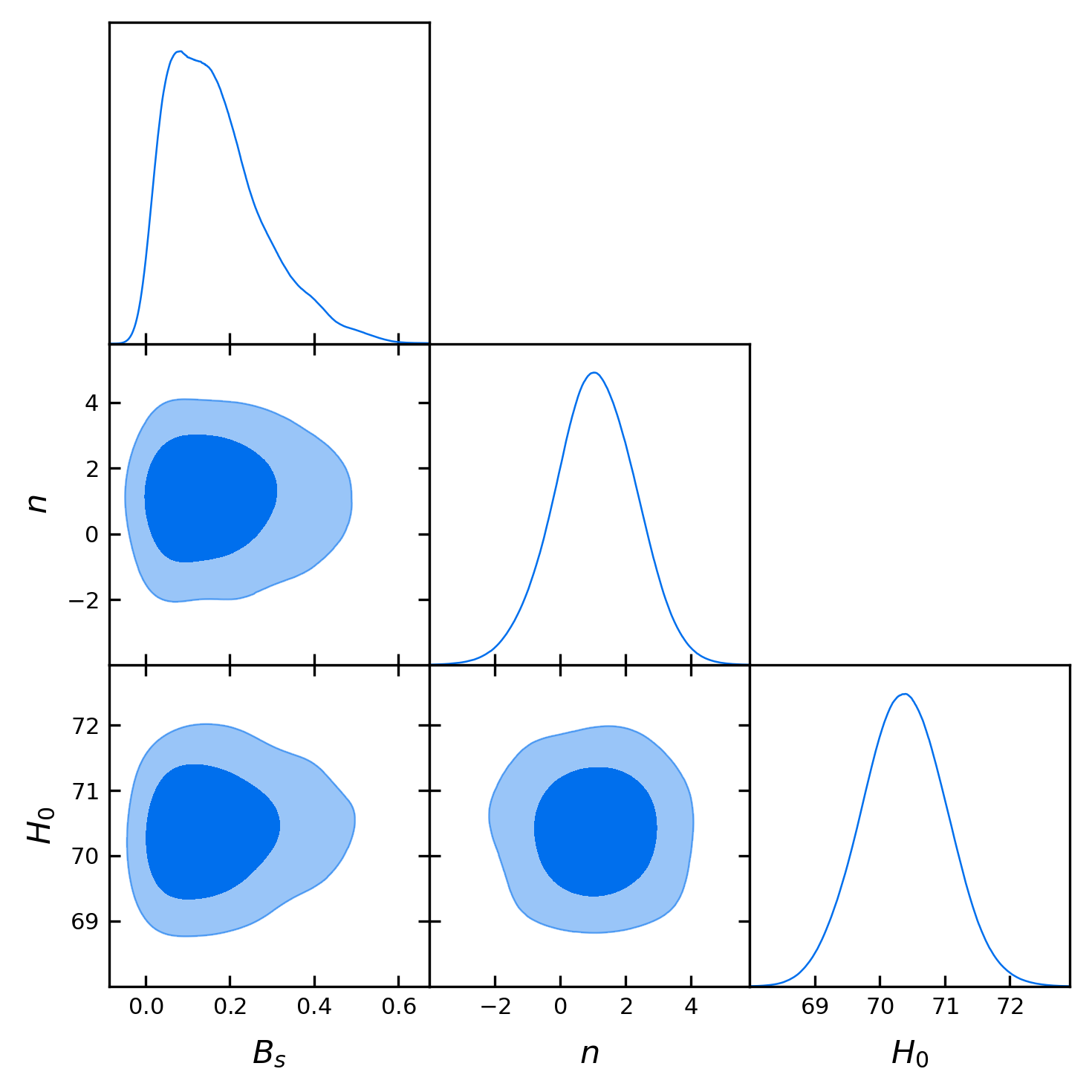}
\caption{Distributions and correlation of constraints obtained on model parameters and $H_0$ from Baryon Acoustic Oscillations data. Results in table (\ref{final_table})}
\label{bao_result}
\end{figure}

\begin{figure}[H]

\includegraphics[width=2in]{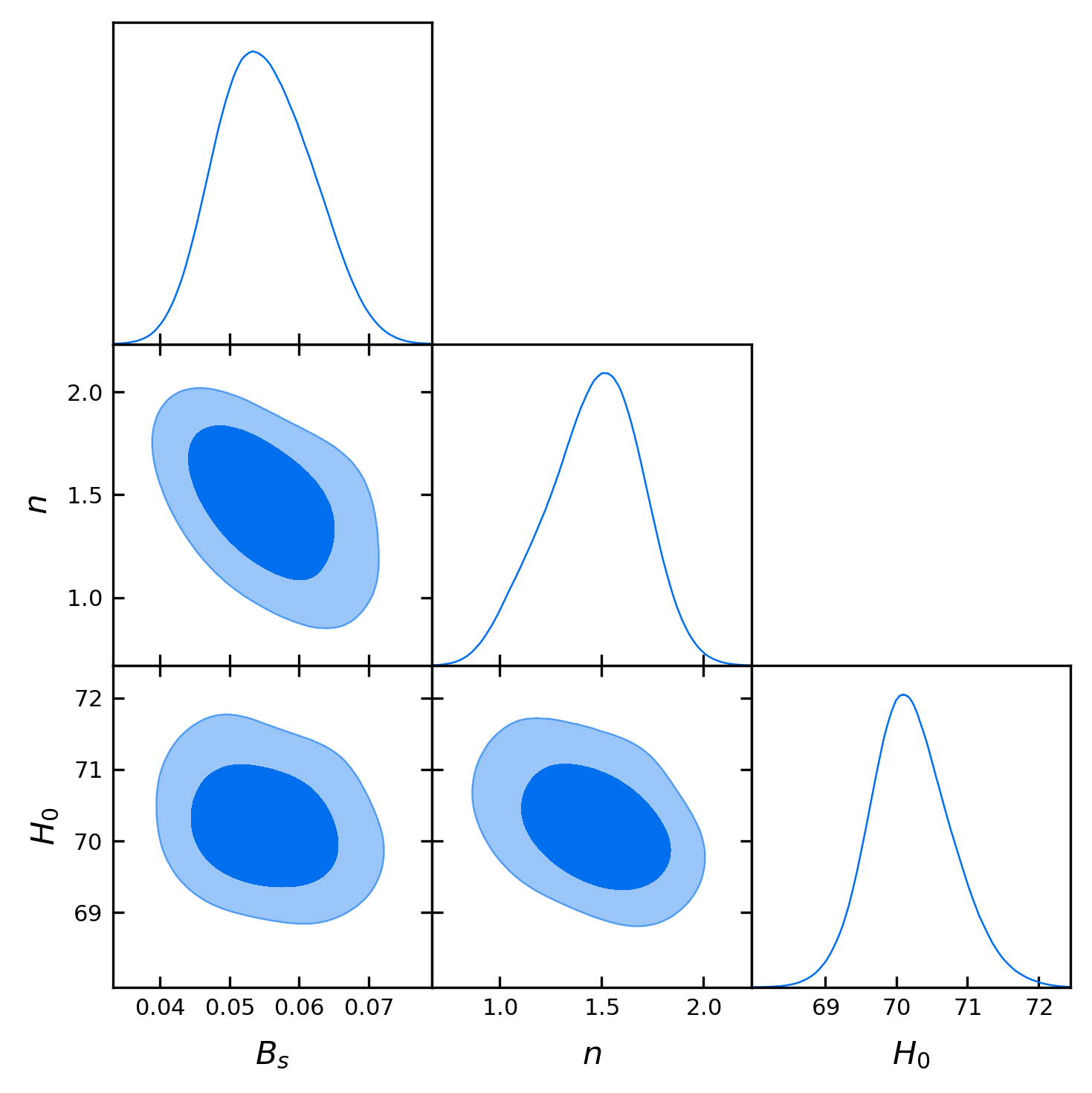}
\caption{Distribution and correlation of constraints obtained on model parameters and $H_0$ from data of Hubble parameter. Results in table (\ref{final_table})}
\label{hz_result}

\end{figure}

\begin{figure}[H]

  \centering

\includegraphics[width=2in]{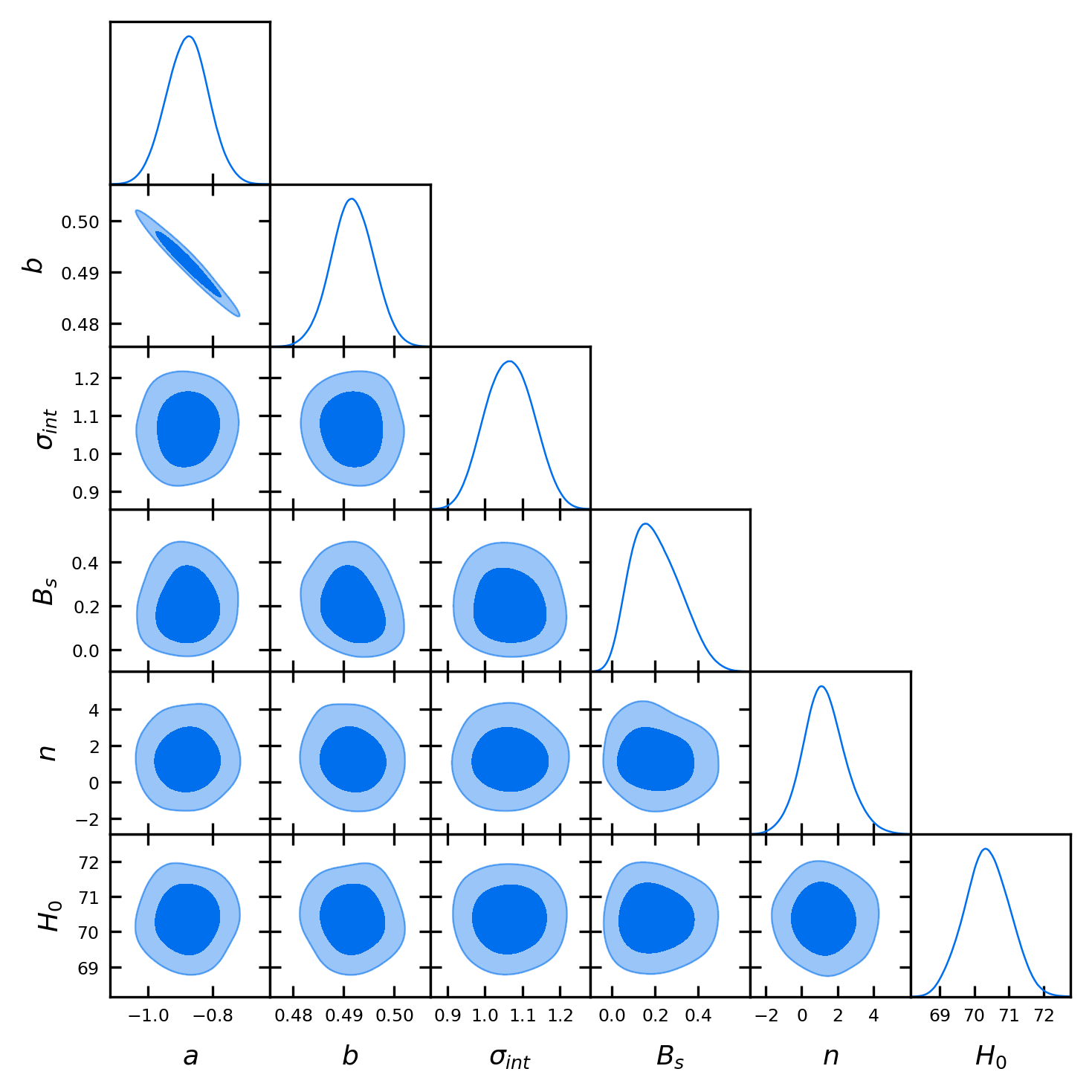}
\caption{Distribution of parameters associated with the modeling of GRB luminosity in equation(\ref{grb_lum}) along with model parameters and $H_0$ for the Platinum GRB sample.Results in table (\ref{final_table}).}

\label{grb_result}
\end{figure}

\begin{figure}[H]
 
\includegraphics[width=2in]{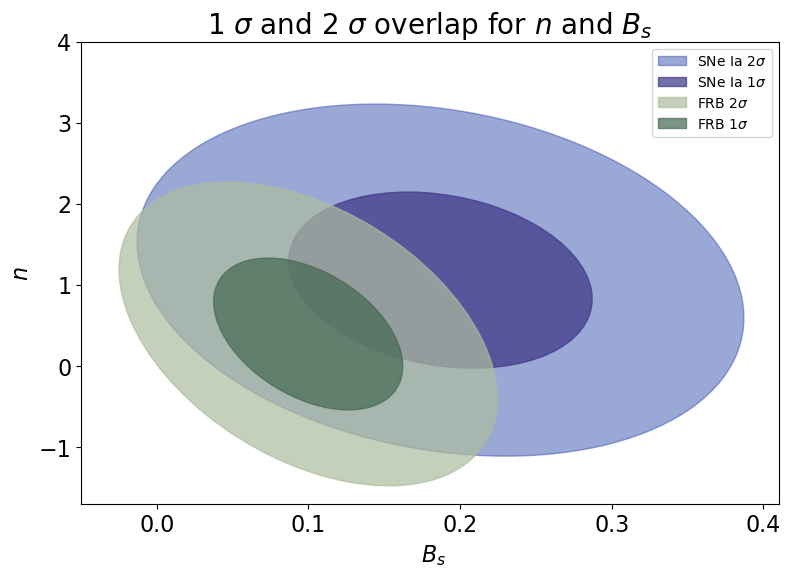}
\caption{ Comparison of contours obtained for model parameters with FRB and Pantheon data.}
\label{sne_frb}
\end{figure}

\begin{figure}[H]
 
\includegraphics[width=2in]{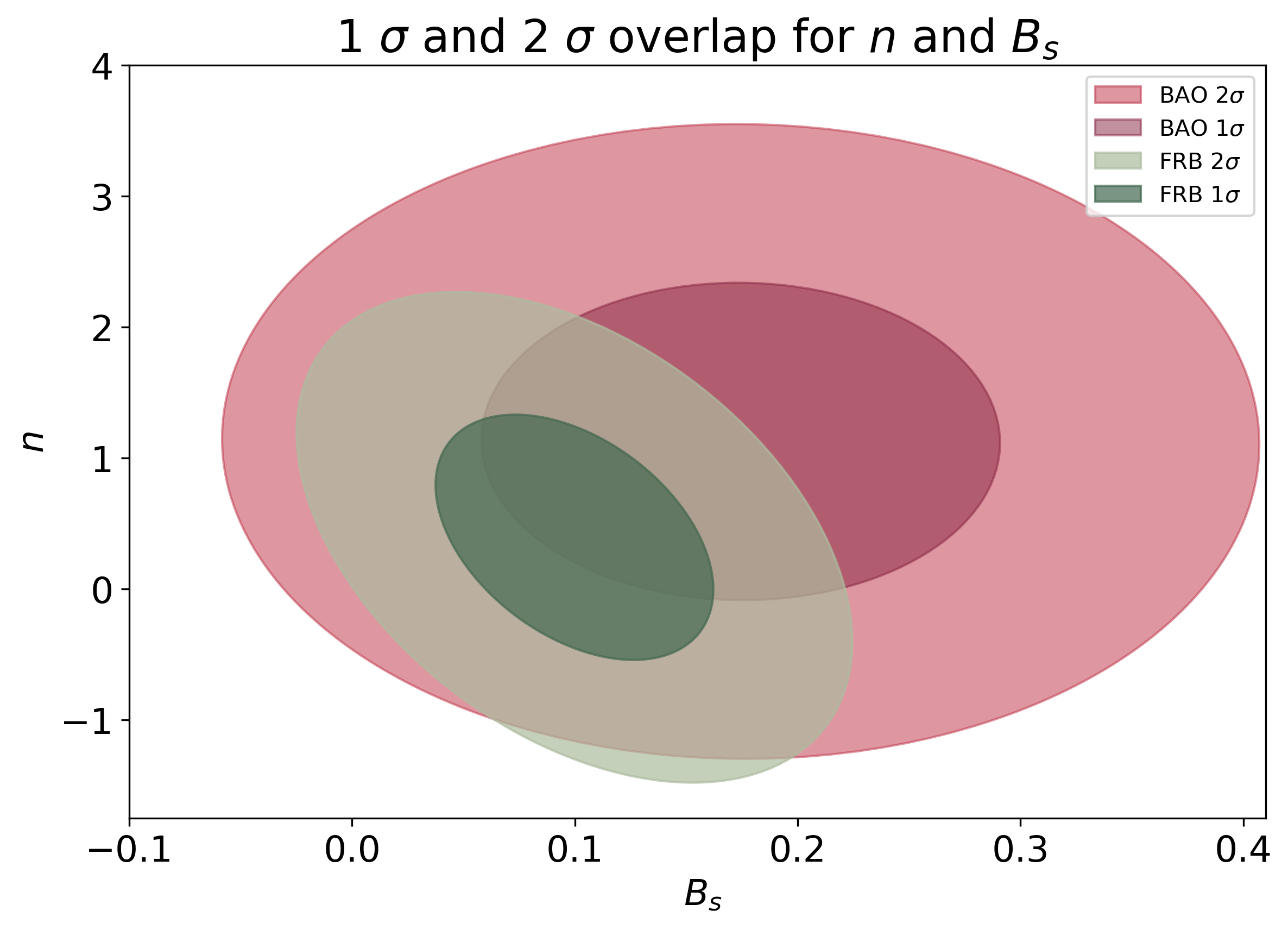}
\caption{ Comparison of contours obtained for model parameters using  Baryon Acoustic Oscillations and FRB data.}
\label{bao_frb}
\end{figure}
\begin{figure}[H]
 
\includegraphics[width=2in]{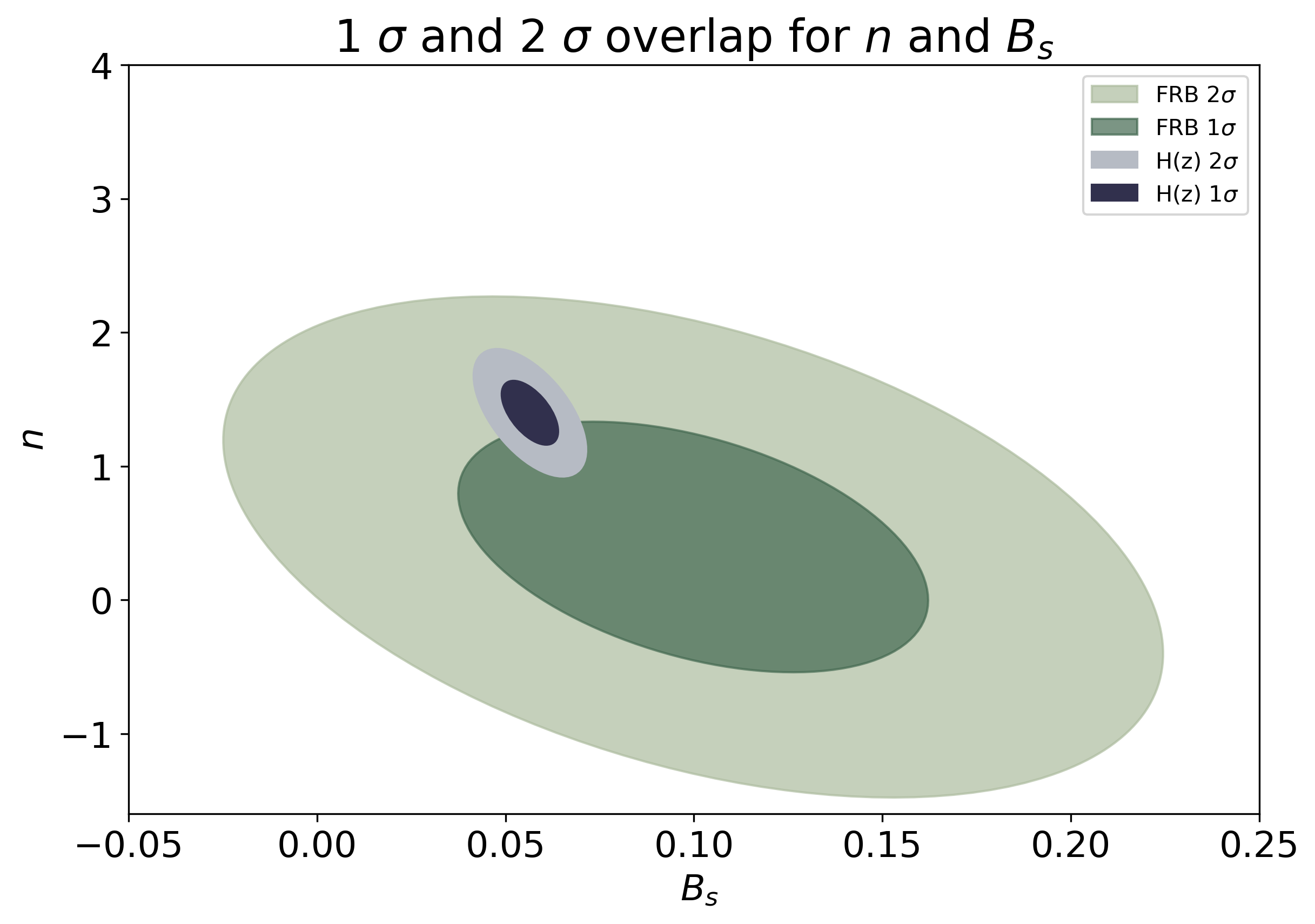}
\caption{Comparison of contours obtained for model parameters  using Hubble parameter data and FRB data.}
\label{hz_frb}
\end{figure}
\begin{figure}[H]
 
\includegraphics[width=2in]{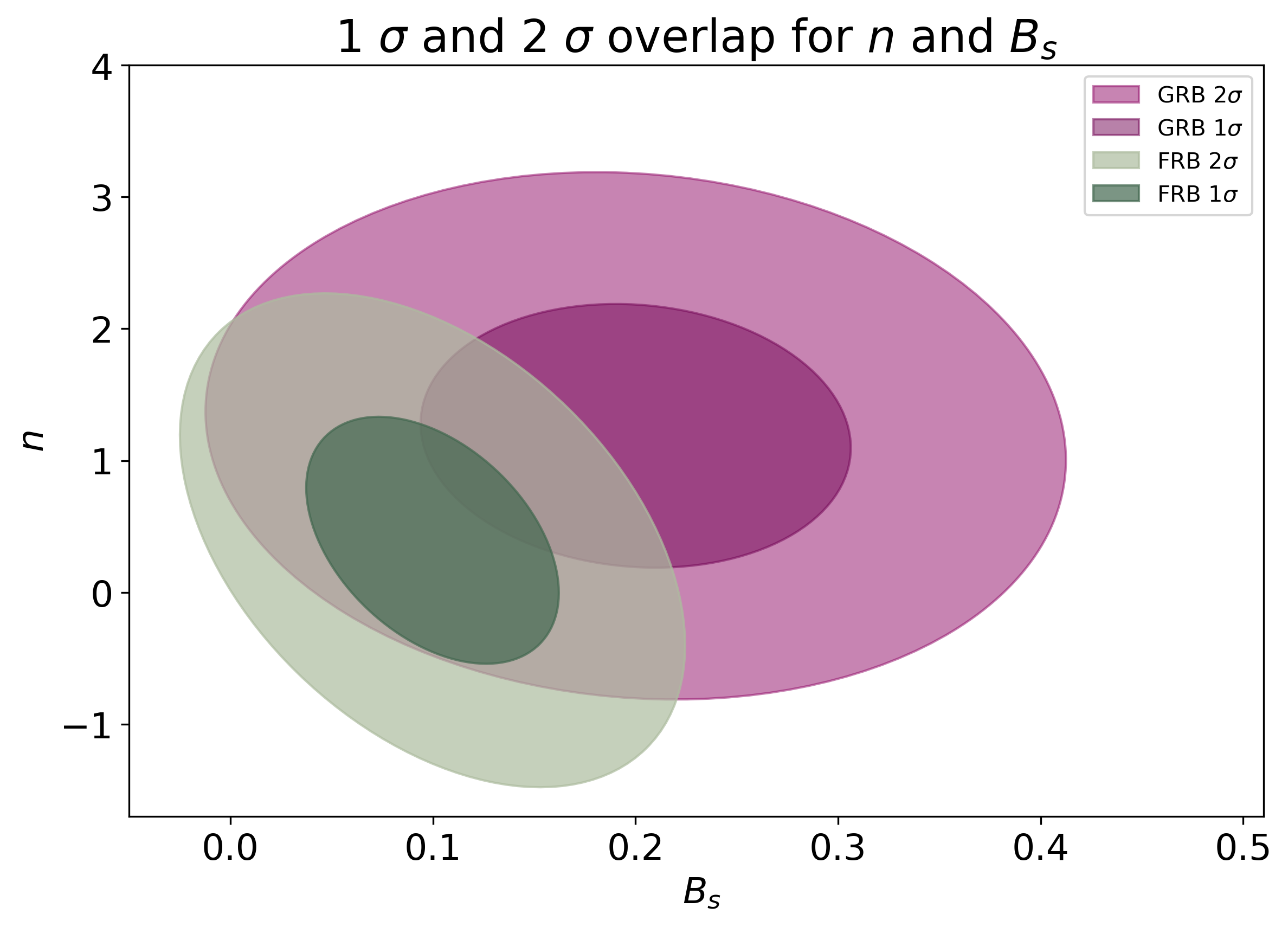}
\caption{ Comparison of contours obtained for model parameters using Platinum GRB data and FRB data.}
\label{grb_frb}
\end{figure}

\begin{comment}

\begin{figure}[H]
\begin{center}

\begin{tabular}{cccc}
\subfloat[Distributions for the parameters $B_s$, $n$ and $H_o$ for $H(z)$ using VCG model]{\includegraphics[width = 2.5in]{fin_hz.png}}\\
\subfloat[ 1 $\sigma$ and 2 $\sigma$ overlaps of $n$ and $B_s$ obtained from SNe Ia and FRB data]{\includegraphics[width = 2.5in]{final_SNe_FRB.png}}\\
\subfloat[ 1 $\sigma$ and 2 $\sigma$ overlaps of $n$ and $B_s$ obtained from BAO and FRB data]{\includegraphics[width = 2.5in]{final_BAO_FRB.png}} 
\end{tabular}

\end{center}
\end{figure}

\begin{figure}[H]
\begin{center}

\begin{tabular}{cccc}

\subfloat[Distributions for the parameters $B_s$, $n$, $H_o$, $\sigma_{int}$ and $b$ for the GRB Data set using VCG model]{\includegraphics[width = 2.5in]{final_grb.png}}\\
\subfloat[ 1 $\sigma$ and 2 $\sigma$ overlaps of $n$ and $B_s$ obtained from H(z) and FRB data]{\includegraphics[width = 2.5in]{final_H_FRB.png}}\\
\subfloat[ 1 $\sigma$ and 2 $\sigma$ overlaps of $n$ and $B_s$ obtained from GRB and FRB data]{\includegraphics[width = 2.5in]{final_GRB_FRB.png}} 
\end{tabular}

\end{center}
\end{figure}

\end{comment}
%%%%%%%%%%%%%%%%%%%%
The results with $1 \sigma$  bounds were shown in Table (\ref{final_table}). It also includes the maximum likelihood corresponding to each dataset. In Table (\ref{Table3}), we have also compared our VCG model results with the GCG model results.

\begin{table*}[htbp]
    \centering
    \begin{tabular*}{\textwidth}{@{\extracolsep\fill}lcccc}
    \hline
    &&&\cr
        Data set & $B_s$ & $n$ & $H_o$ & $\chi^2_{min}(B_s,n,H_0)$ \cr
        &&&\cr
        \hline
        &&&\cr
         Pantheon & $0.18 \pm 0.10$  & $1.10\pm1.15$& $70.46\pm0.66$ & $ 2.82$ \cr
    &&&\cr
    FRB & $0.09\pm0.06$ & $0.44\pm0.89$ &  $70.57\pm0.64$ & $100.65$\cr
    &&&\cr
    Baryon Acoustic Oscillations & $0.16\pm0.11$ & $1.06\pm1.25$ &  $70.37\pm0.65$ & $17026.30$\cr
    &&&\cr
    H(z) data & $0.05\pm0.006$ & $1.46\pm0.23$ &  $70.21\pm0.57$ & $22.97$\cr
    &&&\cr
    Gamma Ray Bursts & $0.20\pm0.11$ & $1.25\pm1.17$ &  $70.37\pm0.64$ & $96.19$\cr
    
     \hline
     &&&\cr

    \end{tabular*}
    \caption{The constrained parameters $B_s$,$n$ and $H_o$ of VCG model obtained from Supernova,FRB, BAO, Hubble, and GRB data}
    \label{final_table}
\end{table*}

The deceleration parameter is defined as, $q=-(\ddot{a}a/\dot{a}^2)$.In general, the expression can be written in terms of the Hubble parameter and redshift as
\begin{equation}
    q(z)=\frac{d \ln H(z)}{d \ln (1+z)}-1 
\end{equation}
where $H(z)$ comes from equation(\ref{eq:friedmanvcg}).\\

\begin{equation}
\begin{split}
\small % Adjust font size
q(z) = & \frac{3\rho_{r0}(1+z)^4+2\rho_{b0}(1+z)^3}{2[\rho_{r0}(1+z)^4+\rho_{b0}(1+z)^3+\rho_{ch0}\{B_s(1+z)^6+(1-B_s)(1+z)^n\}^{1/2}]} \\
& + \frac{\rho_{ch0}\{6B_s(1+z)^6 + n(1-B_s)(1+z)^n\}\{ 2B_s(1+z)^6 +(1-B_s)(1+z)^n\}^{1/2}}{2[\rho_{r0}(1+z)^4+\rho_{b0}(1+z)^3+\rho_{ch0}\{B_s(1+z)^6+(1-B_s)(1+z)^n\}^{1/2}]} \\
& - \frac{\rho_{ch0}\{B_s(1+z)^6+(1-B_s)(1+z)^n\}^{1/2}}{2[\rho_{r0}(1+z)^4+\rho_{b0}(1+z)^3+\rho_{ch0}\{B_s(1+z)^6+(1-B_s)(1+z)^n\}^{1/2}]}
\end{split}
\end{equation}

The parameter of the Equation of State can be written as 
\begin{equation}
    w (z) =\frac{p_\chi(z)}{\rho_\chi(z)}
\end{equation}
For VCG (\cite{makhathini_13}), the equation of state parameter can then be given from equation(\ref{eos_eqn}) as 
\begin{equation}
w = \frac{-V^2 (\phi)}{\rho^2}
\label{w_eos}
\end{equation}

\begin{equation}
\begin{split}
w &=\frac{A_0 a^{-n}}{\rho^2} \\ %I think cite eq for rho here from the paper
w &=\frac{A_0 a^{-n}}{\frac{6}{6-n}B_s(1+z)^6+(1-B_s)(1+z)^n} \\
\end{split} 
\end{equation}

\begin{equation}
    P = -\frac{V_n ^2(\phi)}{\rho}
\end{equation}
Also, we know that 
\[ C^2 _s =\frac{\partial P}{\partial \rho}    \]
from equation(\ref{w_eos}), we can write the square of the speed of sound as 
\begin{equation}
C^2 _s = \frac{V^2 (\phi)}{\rho^2}
\end{equation}

\begin{equation}
\begin{split}
C^2 _s &=\frac{-A_0 a^{-n}}{\rho^2} \\ %I think cite eq for rho here from the paper
C^2 _s &=\frac{-A_0 a^{-n}}{\frac{6}{6-n}B_s(1+z)^6+(1-B_s)(1+z)^n} \\
\end{split}   
\end{equation}
\vspace{5cm}
\begin{figure}[H]
 \flushright
\includegraphics[width=2.5in]{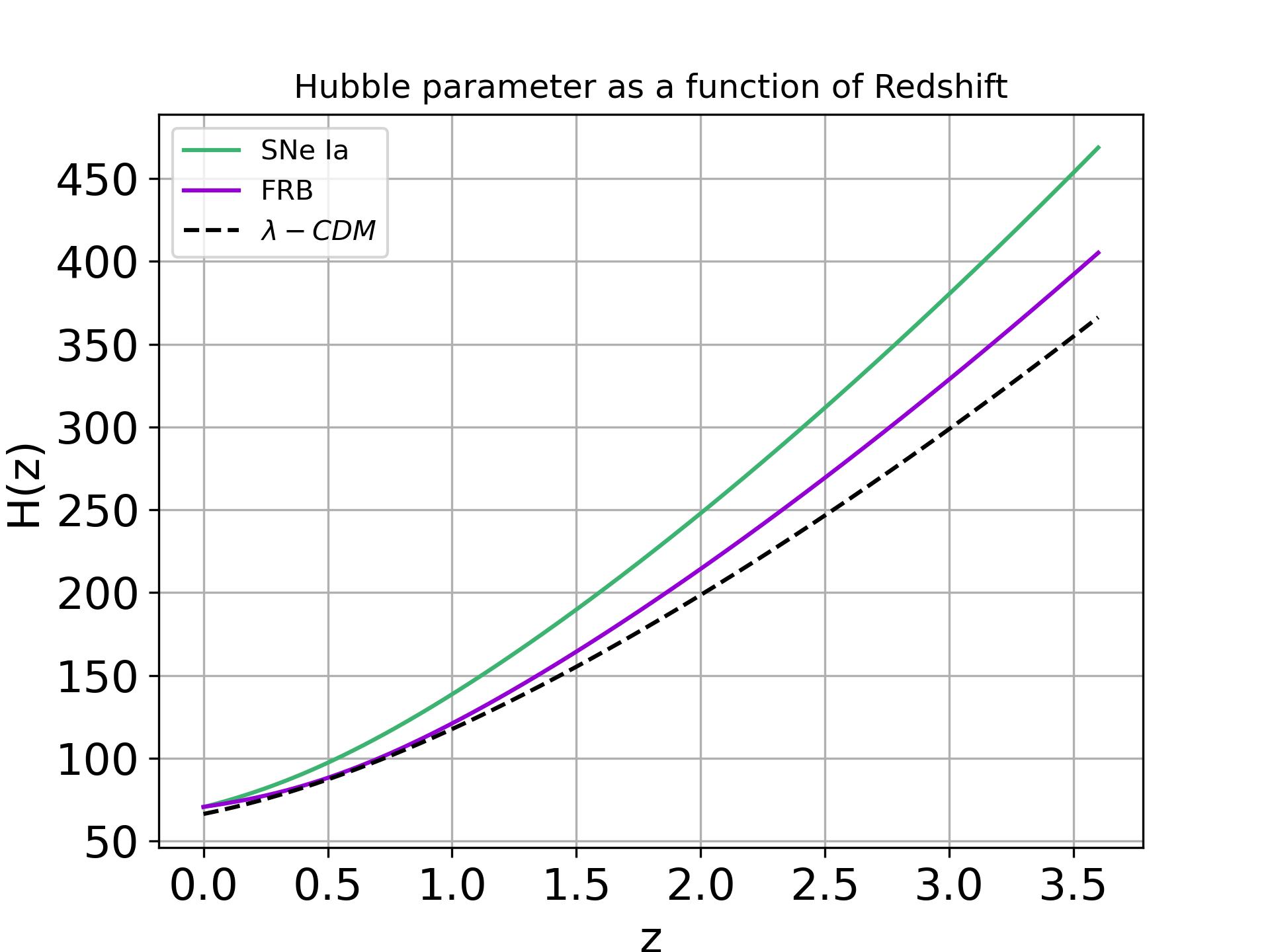}
\caption{Evolution of $H(z)$ with redshift}
\end{figure}
\begin{figure}[H]
\flushright
\includegraphics[width=2.5in]{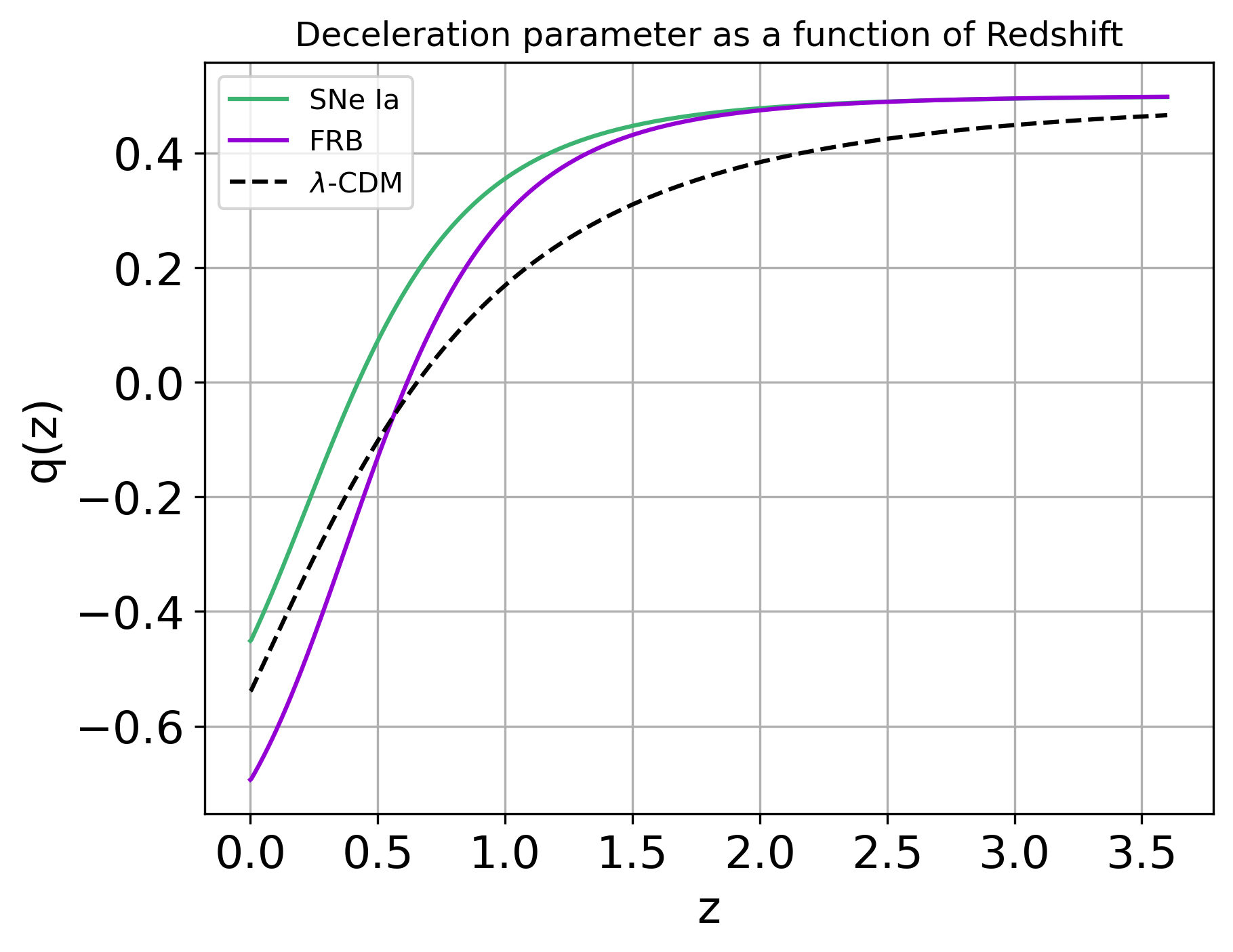}
\flushright
\caption{Evolution of deceleration parameter $q(z)$ with redshift }
\end{figure}

\begin{figure}[H]
\flushright
\includegraphics[width=2.5in]{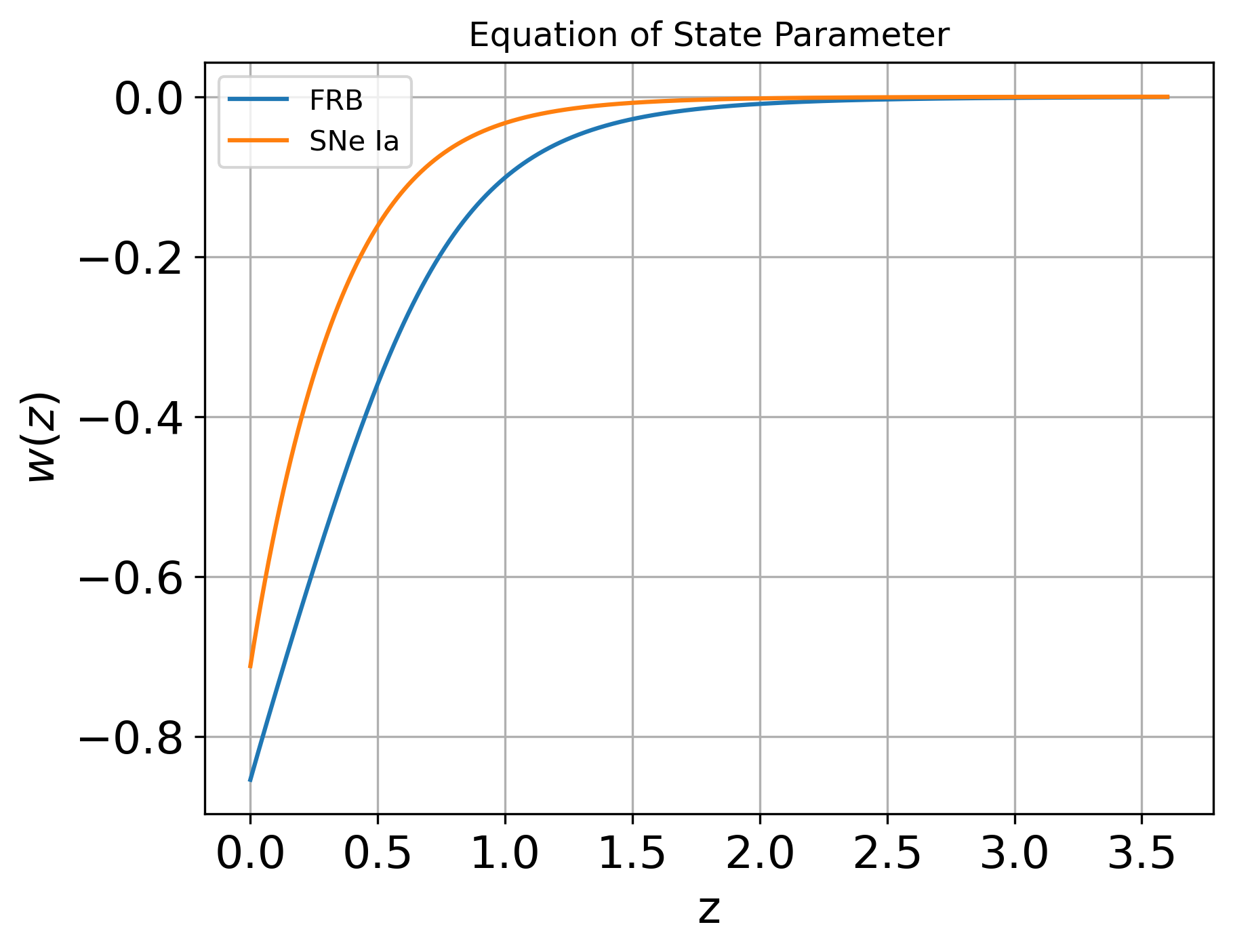}
\caption{Evolution of Equation of state parameter $w(z)$ with redshift}
\end{figure}
\begin{figure}[H]
\flushright
\includegraphics[width=2.5in]{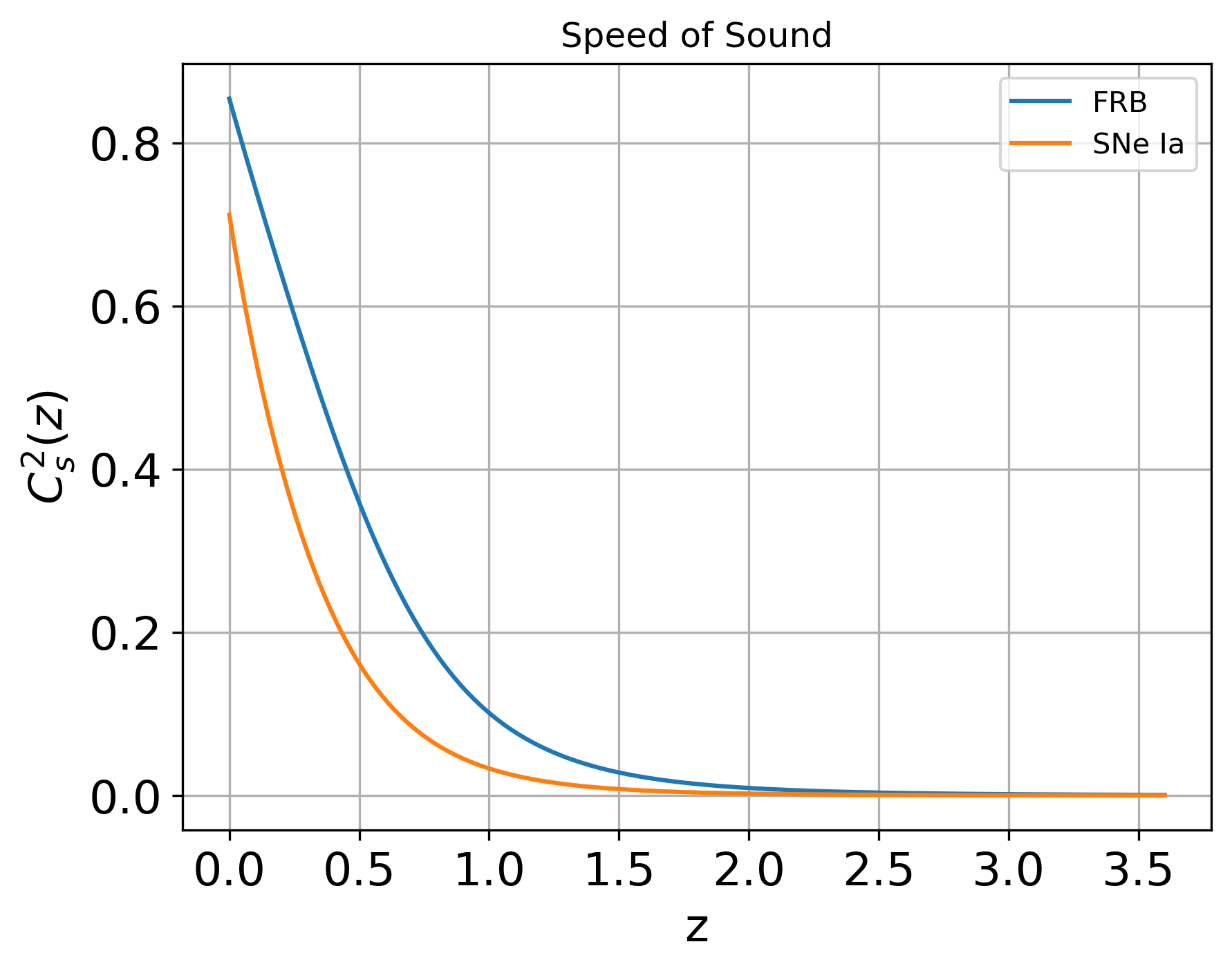}
\caption{Evolution of speed of sound  $C^2 _s$ with redshift}
\end{figure}
 These parameters are plotted in Fig (10) to Fig (13) for best-fit values of $B_s$ and n obtained from the FRB and SNe Ia analysis. The present values of these parameters as obtained using FRB results are $H(0)=70.6$, $w(0)=-0.85$, $c_s^2(0)=0.85$ and $q(0)=-0.69$. The redshift of transition always lies between $1.5$ to $2.0$.

\section{Dynamical stability of the model with results from FRB}
In this section we have discussed the dynamical stability of the Variable Chaplygin Gas for the best-fit values of our parameters obtained in the previous section.

 On converting the physical parameters in the equation of state of the VCG model equation (\ref{chaplygineq}) to dimensionless parameters,
\begin{equation}
 \begin{aligned}
 x&=\ln a \\
 u&=\Omega_{VCG}=\frac{\rho_{VCG}}{3H^2}\\
 v&=\frac{p_{VCG}}{3H^2}\\
\end{aligned}
\end{equation}

The equation of state for VCG then becomes\\
\begin{equation}
    \omega_{VCG}(x)=\frac{p_{VCG}}{\rho_{VCG}}=\frac{v}{u}
\end{equation}
Dimensionless density parameter for barotropic fluid: $\Omega_{\gamma}=\dfrac{\rho_{\gamma}}{3H^2}$ and Friedmann equations are used to obtain\\
\begin{equation}
    \Omega_{\gamma}=1-\Omega_{VCG}=1-u
\end{equation}
Here $\rho_{\gamma}$ is the density of the linear barotropic fluid.
Since, the energies  cannot be negative, for a flat universe $0\leq u\leq 1$\\
The evolution equations of the autonomous systems of $u$ and $v$ can be written as
%\begin{equation}
%    \begin{aligned}
%        \frac{du}{dx} &= -3c-3(u-1)(u\omega_{dm}-v) \\
 %       \frac{dv}{dx} &= \left[A(1+\alpha)-\alpha\frac{v}{u}\right]\left[-3c-3 \left(1+\frac{v}{u}\right)u\right] \\ 
 %       &-n\left(A-\frac{v}{u}\right)u \\
  %      &+3v\left[\left( 1+\frac{v}{u}\right)u+(1+\omega_{dm})(1-u) \right]
 %   \end{aligned}
%\end{equation}
%which for the values of $\alpha$ and $A$ for VCG model,\\

\begin{equation}
\label{stability_eqn_1}
    \begin{aligned}
        \frac{du}{dx} &= -3c-3(u-1)(u\omega_{\gamma}-v) \\
        \frac{dv}{dx} &= \left[-\frac{v}{u}\right]\left[-3c-3 \left(1+\frac{v}{u}\right)u\right] \\ 
        &-n\frac{v}{u}u+3v\left[\left( 1+\frac{v}{u}\right)u +(1+\omega_{\gamma})(1-u) \right]
    \end{aligned}
\end{equation}
where,\\

$c=2u(u-1)$and $\omega_{\gamma}= \frac{p_{\gamma}}{\rho_{\gamma}}$. 
The critical point of the above system of equations are the solutions of the equations $\frac{du}{dx}=\frac{dv}{dx}=0 $ which are obtained to be,\\
\begin{equation}
\label{stability_eqn_2}
    \begin{aligned}
        u_{crit} &= \frac{2(n-3)(1-c+\omega_{\gamma})}{2(n-3)(1+\omega_{\gamma})}\\
        v_{crit} &= \frac{n^2-6n(2+\omega_{\gamma})+36(1+\omega_{\gamma}+c\omega_{\gamma})}{6[6(1+\omega_{\gamma})-n]}
    \end{aligned}
\end{equation}
For a spatially flat universe, the physical meaningful range of $u$ is $0\leq u \leq 1$, and hence, $0 \leq u_{crit}\leq 1$ which gives the equation,\\
\begin{equation}
    0<c\leq \frac{6(1+\omega_{\gamma})-n}{6}
\end{equation}
and a condition for $n$,\\
\begin{equation}
    n\leq min(6(1+\omega_{\gamma}),4)
\end{equation}
The condition of c represents an energy transfer from dark matter to VCG.
To find the stability around the critical point, we attempt to linearize the equations around the critical points, at $u=u_{crit}+\delta u$ and $v=v_{crit}+\delta v$ \\
\begin{equation}
    \begin{aligned}
        \delta\left(\frac{du}{dx}\right)&=[3(v+\omega_{\gamma}+2u\omega_{\gamma})]_{crit}\delta u \\
        &+ [3(-1+u)]_{crit}\delta v \\
        \delta\left(\frac{dv}{dx}\right)&= [-3\omega_{\gamma}v-]_{crit} \delta v +\\
        &\left[(3-n+6v+3(1-u)\omega_{\gamma}){u}\right]_{crit}\delta v
    \end{aligned}
\end{equation}
Eigenvalue equation,\\
\begin{equation}
\begin{aligned}
&\lambda_{1,2} = \frac{1}{2} \left[ 3n - 6(3+c+\omega_{\gamma}) + \frac{36c^2}{6c+n-6(1+\omega_{\gamma})} \right]  \\
& \pm \frac{1}{4} \left\{ \frac{(6-n)^2 + 72c \omega_{\gamma} - 36 \omega_{\gamma}^2}{6c+n-6(1+\omega_{\gamma})} \right\}^{1/2} \times \\
& \frac{\left\{-144c^2 + (6-n)^2 - 36 \omega_{\gamma}^2 - 24c(n-6-9\omega_{\gamma})\right\}^{1/2}}{6c+6(1 + \omega_{\gamma})}
\end{aligned}
\end{equation}

\begin{figure}[H]
 
\includegraphics[width=2.5in]{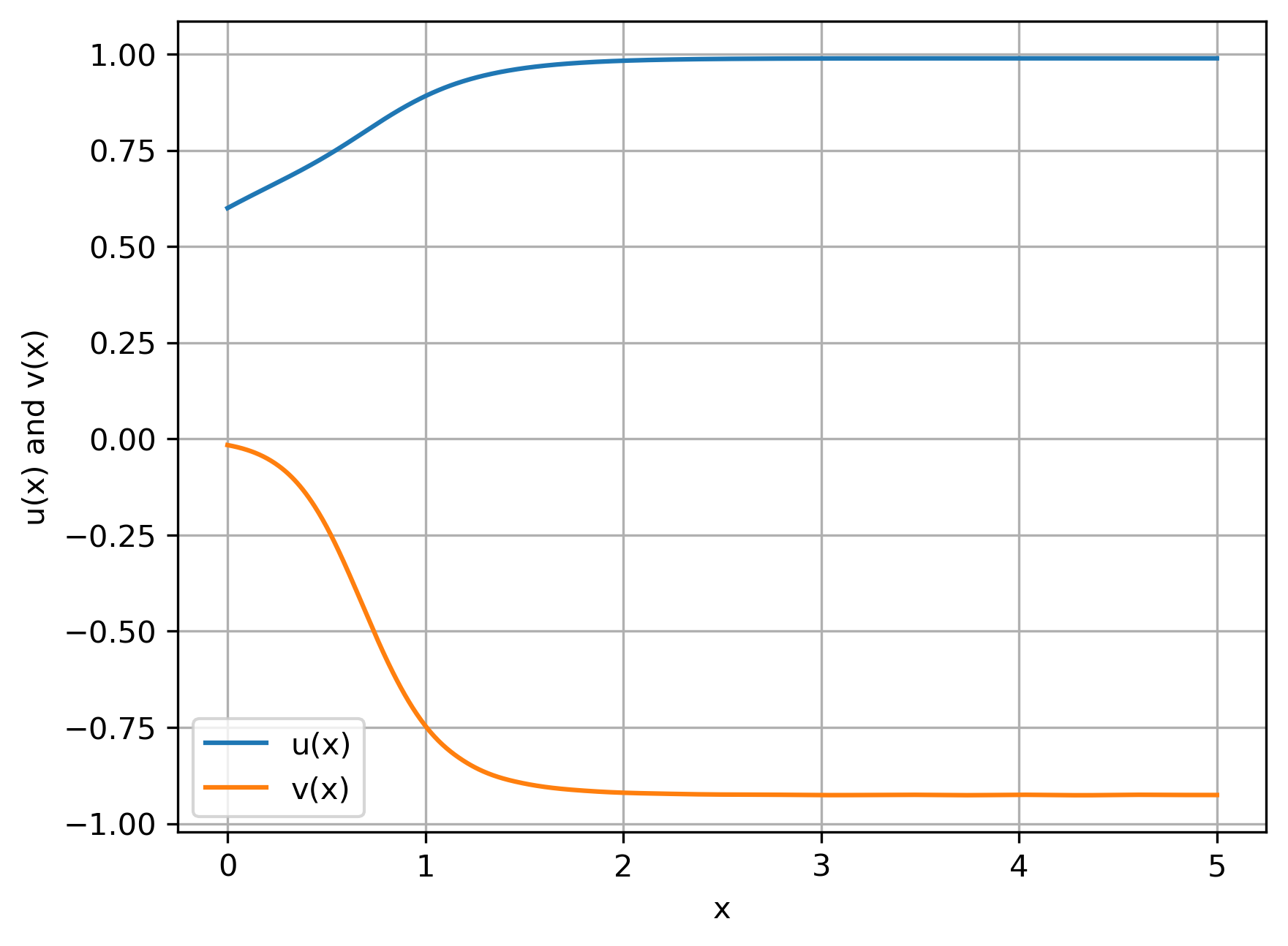}
\caption{Evolution of u and v as a function of z}

\end{figure}
\begin{figure}[H]
\includegraphics[width=2.5in]{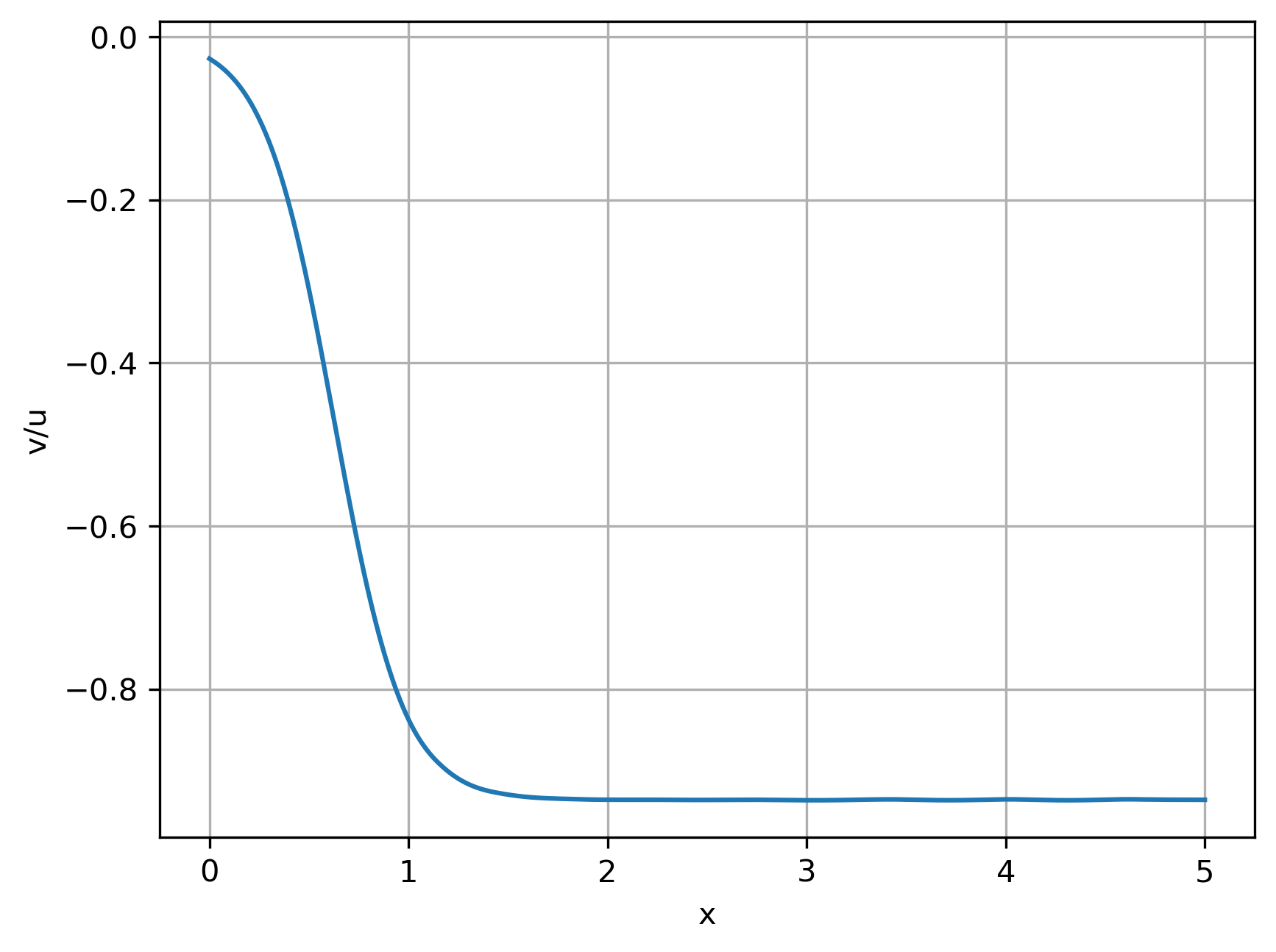}
\caption{$v/u$ as a function of z}
\end{figure}

If the real parts of the above eigenvalues are negative, the critical point is a stable node and is a stationary
attractor solution; otherwise unstable and thus oscillatory. The physical meaningful range of c is $0 \le c \le \frac{6(1+\omega_{\gamma})-n}{6}$ and in this range the critical point $(u_{crit},v_{crit})$ is stable and is a late-time stationary attractor solution. Here we plot  figures to show the properties of the evolution of the Universe controlled by the dynamical system (32). The dimensionless parameters u and v have been drawn in Fig (14) and Fig (15)
in terms of $x = ln a$ for our best-fit parameters.
\begin{table}[htbp]
    \small % Adjust the font size to make the table smaller
    
    \begin{adjustbox}{width=1.2\columnwidth, center} % Adjust the width as needed
        \setlength{\tabcolsep}{4pt} % Adjust column separation
        \renewcommand{\arraystretch}{1.2} % Adjust row height
        
        \begin{tabular}{lcccc}
            \toprule
            Dataset & $h$ & $\Omega_{b0}h^2$ & $B_s$ & $n$ \\
            \midrule
            \multicolumn{5}{c}{VGCG} \\
            \midrule
            SNe Ia & $0.70\pm0.66$ & $0.023\pm0.021$ & $0.18\pm0.10$ & $1.10\pm1.15$ \\
            FRB & $0.71\pm0.64$ & $0.024\pm0.020$ & $0.09\pm0.06$ & $0.44\pm0.89$ \\
            \midrule
            \multicolumn{5}{c}{GCG} \\
            \midrule
            SNe Ia & $0.71^{+0.20}_{-0.18}$ & $0.02242^{+0.00014}_{-0.00014}$ & $0.96^{+0.05}_{-0.04}$ & $1.13^{+0.44}_{-0.28}$ \\
            \bottomrule
        \end{tabular}%
    \end{adjustbox}
    
    \caption{Results of constraining the cosmological parameters with Pantheon dataset in other dissipative fluids models of cosmology like VGCG and GCG (\cite{HER-AL,refId01,FREITAS2011209})}
    
    \label{Table3}
\end{table}

\section{Conclusion}
The Variable Chaplygin gas model is indeed able to explain the accelerated evolution of the universe. As the VCG Model speculates, the Chaplygin gas is expected to behave like a non-relativistic entity and later evolve to account for the accelerated expansion observed in the current epoch of the Universe. The best-fit values obtained for the VCG model with the FRB data sets lie within confidence levels obtained in previously conducted analysis  (\cite{Sethi_2006}) and (\cite{Chraya_2023}) using various cosmological probes. Although the value of $H_0$ obtained using FRB data set is in good agreement with value obtained using the SNe Ia data, the Variable Chaplygin Gas model is still not able to resolve the Hubble Tension problem. One can try to obtain constraints using the CMB data and better and larger FRB data in the future. We have also plotted the behaviour of $H(z)$, $q(z)$, $w(z)$ and $c_s^2(z)$ for the best fit values of the parameters. The system is dynamically stable for our best-fit values.

\section{Declarations}
\textbf{Competing Interests} The authors declare no competing interests. 

\section{Acknowledgments}
The authors are grateful to the Principal, St. Stephen's College for his support and encouragement towards this research.  We also thank Dr. Akshay Rana, St. Stephen's College and Mr. Ashley Chraya, Vanderbilt University for their input and suggestions. We are also grateful to the reviewers for their suggestions and input.

\bibliography{references}
\newpage

\appendix

\section{Appendix: Stability Analysis}
Derivations for equations (\ref{stability_eqn_1}) and (\ref{stability_eqn_2})
\begin{equation*}
\begin{split}
u&=\frac{k^2 \rho_{VCG}}{3H^2} ; \ \ v= \frac{k^2 p_{VCG} }{3H^2} ; \ \ N=ln(a) ;\\ \omega_{\gamma}&= \frac{p_{\gamma}}{\rho_{\gamma}} ; \ \ \dot{\rho_{VCG}} = -3H(\rho_{VCG} + p_{VCG})\\
\frac{dN}{dt} &= \frac{\dot{a}}{a} =H \\
\frac{du}{dN} &= \frac{du}{dt}\frac{dt}{dN} = \frac{1}{H}\frac{du}{dt} \\
\end{split}
\end{equation*}
Here t is the cosmic time.
We can write
\begin{equation*}
\begin{split}
1&= \frac{k^2}{3H^2} \rho_{\gamma} + \frac{k^2}{3H^2} \rho_{v}\\
\frac{k^2}{3H^2} \rho_{\gamma}&= 1 - \frac{k^2}{3H^2} \rho_{v} 
\end{split}
\end{equation*}
\begin{equation*}
\begin{split}
\frac{du}{dt} &= \frac{k^2}{3H^2} \dot{\rho_{VCG}} + \frac{k^2 \rho_{VCG}}{3} \frac{d}{dt} (H^{-2}) \\
&= \frac{k^2}{3H^2}(-3H)(\rho_{VCG} + p_{VCG}) + \frac{k^2 \rho_{VCG}}{3} (-2) H^{-3} \dot{H}\\
&= \frac{k^2}{3H^2}(-3H)(\rho_{VCG} + p_{VCG})  -2\frac{k^2 \rho_{VCG}}{3H^2}  \frac{\dot{H}}{H}
\end{split}
\end{equation*}
\begin{equation*}
\begin{split}
&\frac{du}{dN} = \frac{1}{H} \frac{du}{dt} =-3 \Big[ (u+v)  
\\ &-\frac{u}{H^2}\Big\{-k^2(\rho_{\gamma}+ p_{\gamma} + \rho_{VCG} + p_{VCG} \Big\} \Big] \\
&=-3(u+v) + 3u \Big\{ \frac{k^2}{3H^2}(\rho_{\gamma}+ p_{\gamma} + \rho_{VCG} + p_{VCG}) \Big\} \\
&=-3(u+v) + 3u \Big\{ \frac{k^2}{3H^2}(\omega_{\gamma} + 1)\rho_{\gamma} + (u+v) \Big\}\\
&=-3(u+v) + 3u \Big\{ (\omega_{\gamma} + 1)(1-u) + (u+v) \Big\}  \\
&=-3(u+v) + 3u \omega_{\gamma}(1-u) + 3u(u-1) + 3u(u+v) \\
&=-3(u+v) + 3u \omega_{\gamma}(1-u) +3u^2 -3u + 3u^2 + 3uv \\
&= -3c - 3(u-1)u\omega_{\gamma} + 3(u-1)v 
\end{split}
\end{equation*}

\end{document}